\documentclass[showpacs, pra,onecolumn,preprintnumbers ,amsmath, amssymb, superscriptaddress, aps]{revtex4-2}
\usepackage{color}
\usepackage{amsmath,amssymb}
\usepackage{pifont}% http://ctan.org/pkg/pifont
\usepackage{amssymb}  % More symbols
\usepackage{bbold}
\usepackage{float}
\usepackage{subfloat}

\usepackage[caption=false]{subfig}
\usepackage{tikz}
\usepackage{makecell}
\usepackage{subfig}
\usepackage{pifont}   % Ding symbols
\usepackage{graphicx} % Include figure files
\graphicspath{{Figures/}}
\usepackage{dcolumn}  % Align table columns on decimal point
\usepackage{bm}       % bold math
\usepackage{multirow} % Table functions
\usepackage{placeins}% For making floats not move around everywhere
\usepackage[colorlinks]{hyperref}
\usepackage{mathtools}

\begin{document}

	\title{ Effect of  magnetic field on  fermions in graphene through time-oscillating potential  }
	\date{\today}
	\author{Rachid El Aitouni}
	%\email{}
	\affiliation{Laboratory of Theoretical Physics, Faculty of Sciences, Choua\"ib Doukkali University, PO Box 20, 24000 El Jadida, Morocco}
	\author{Ahmed Jellal}
	\email{a.jellal@ucd.ac.ma}
	\affiliation{Laboratory of Theoretical Physics, Faculty of Sciences, Choua\"ib Doukkali University, PO Box 20, 24000 El Jadida, Morocco}
	\affiliation{Canadian Quantum Research Center, 204-3002 32 Ave Vernon,  BC V1T 2L7, Canada}


	\begin{abstract}	
We study the effect of a magnetic field on  Dirac fermions in graphene subject to a scalar potential oscillating in time. Using the  Floquet theory and resonance approximation, we show that the energy spectrum exhibits   extra subbands resulted from the  oscillating potential in addition to quantized Landau levels. It is found that a current density can be generated in $ x $ and $ y $-directions that is strongly dependent on  the magnetic field and  potential. Our numerical analysis show that  the energy spectrum possesses a symmetry and the current density oscillates with different amplitudes under various conditions.
		
	\end{abstract}

	\maketitle

	\section{Introduction}
	Graphene is one of the most two-dimensional materials that received a huge study starting from the first discovery in 2004 \cite{2}. 
	This comes because such a material possesses incredible properties, either electric, mechanical, or optics \cite{9}. Just to mention,   electrons in graphene  have a very great speed  and behave like massless Dirac fermions. Graphene has a linear dispersion relation in the energy spectrum  \cite{3} and  high mobility \cite{4}. 
	However, the absence of a gap 
	  causes a delay in finding suitable applications for graphene in the industry.
	 It turns out, however 
	 a gap is needed to control the flow of the charge carriers in   systems based on graphene.
	 This Dirac gap can be opened using a variety of experimental methods \cite{300}.
	 Due to the sublattice symmetry breakdown, the greatest energy gap might be 260 meV, as proven in the experiment \cite{250}.
	 It's worth noting that changing the experimental technique changes the value of the energy gap.
	 Controlling the structure of the graphene-ruthenium interface has been established as one of the experimental ways of opening a gap \cite{260}.
	 Furthermore, an energy gap has been detected in graphene grown epitaxially on a SiC substrate \cite{250}. In addition,
	it has been theoretically established that an energy gap can be closed using a local strain and/or a chemical approach \cite{270,280,290}.
	
	On the other hand, quantum transport in periodic-driven systems is a crucial topic for device and optical applications as well as for academic purposes.
	Extra sidebands in the transmission probability are caused by electrons exchanging energy quanta carried by the oscillating field, as described by
	\cite{Sabeeh08,Mekkaoui2014}.
With this respect, the standard model is a time-modulated scalar potential in a finite region of space, studied experimentally \cite{444}   by showing evidence of photon-assisted tunneling in superconducting films under microwave fields. 
Theoretically,  the first explanation for these experimental findings was reported in \cite{555}.
	The recent upsurge in theoretical analysis of the influence of 
	time-dependent periodic electromagnetic fields on electron spectra was driven by the growing experimental interest in studying optical features of electron transport in graphene under  intense laser beams \cite{1100}.
	Laser beams have recently been shown to change the electron density of states and, as a result, electron transport properties \cite{1200}.
	Subharmonic resonant enhancement was seen in graphene after electron transport was induced by laser irradiation \cite{1300}. 
	Recent research \cite{1700} has verified that an applied oscillating field can result in an effective mass or, in other words, a dynamic gap.
	A Josephson-like current was anticipated for numerous time-dependent scalar potential barriers placed on a monolayer of graphene
	\cite{1900}. 
It is shown that
	the electron spectrum of graphene superlattices created by static one-dimensional periodic potentials is analogous to the spectra of Dirac fermions in laser fields \cite{5}.

	We study the effects of a magnetic field on Dirac fermions in graphene, subjected to oscillating potential in time. The solutions of the energy spectrum are found by using the Floquet theory in the first order approximation together with the algebraic method, which results in an energy with two subbands. 
	 We determine the current density and in $x$ and $y$-directions, show the contribution of the oscillating potential. We numerically analyze our results and  in particular show that 
	the current density oscillates with different amplitudes depending on the magnetic field and the  potential.

The paper  is organized as follows. In  section \ref{EnergyS}, we establish the theoretical model describing Dirac fermions in graphene in the presence of a magnetic field and oscillating potential. 
The algebraic method is employed to find the eigenvalues and eigenspinors  for the Hamiltonian part time-independent, resulting in Landau levels.
We completely determine the solutions of the energy spectrum by using the Floquet theory in 
section \ref{LLs}. 
 The numerical analysis shows that the quantized energy is shifted due to the subbands resulting from the oscillating potential.   The current density and the graphical representations under various conditions of the physical parameters  will be the subject of section  \ref{CurrentD}. Finally, we conclude our work.

	\section{Landau levels	}\label{EnergyS}
	
	In this paper, we consider graphene in the presence of a magnetic field $B$ and a periodic scalar potential with amplitude $U_0$ and frequency $w$ in time. 
	The Hamiltonian (in the unit $\hbar=e=v_F=1$) describes this system as two parts
	\begin{equation}\label{1}
		H=H_0+H_1
	\end{equation}
	such that  $H_0$ reads as
	\begin{eqnarray}
		H_0=\vec \sigma \cdot (\vec p-\vec A)
	\end{eqnarray}
and  the time-dependent Hamiltonian is given by
	\begin{eqnarray}
		H_1=U_0\cos(wt)\mathbb{I}_2
	\end{eqnarray}
	where  $\sigma_{i}$ ($i=x,y$) are the Pauli matrices, 
	$ \vec p-\vec A $ is the conjugate momentum chosen in the Landau gauge of 
	  the vector potential   $\vec{A}=(0,Bx,0)$ and $\mathbb{I}_2$ is the unit matrix. 
	To explicitly determine the solutions of the energy spectrum, we separately treat  each part of the Hamiltonian \eqref{1}.
	Because momentum $p_y$ is a conserved quantity in the Hamiltonian $H_0$, the corresponding eigenspinors can be 
		$\psi(x,y)=\left[\psi_A(x),\psi_B(x)\right]^T e^{ik_yy}$ and
	distinguished by the labels "A" and "B" on the two triangular sublattices. We continue by solving the eigenvalue equation $H_0\psi(x,y) = \varepsilon \psi(x,y) $ related to the energy $\varepsilon $. It can be written as
	\begin{align}\label{5}
		\begin{pmatrix}
			-\varepsilon & p_x-i(k_y-Bx)\\
			p_x+i(k_y-Bx) & -\varepsilon
		\end{pmatrix}
		\begin{pmatrix}
			\psi_A\\
			\psi_B
		\end{pmatrix}
		=\begin{pmatrix}
			0\\
			0
		\end{pmatrix}.
	\end{align}
	%\end{widetext}
To go further, 	it is convenient to   apply the algebraic method based on the annihilation and creation operators. They can be mapped as
	\begin{align}
&	a=\frac{1}{\sqrt{2B}}\left[p_x+i(k_y-Bx)\right]\\
	&	 a^\dagger = \frac{1}{\sqrt{2B}}\left[p_x-i(k _y-Bx)\right]
	\end{align}
	which fulfill  the commutation relation $[a,a^\dagger]=\mathbb{I}$. These operators can be used to write
	\eqref{5} as
	\begin{equation}
		\begin{pmatrix}
			-\varepsilon & \sqrt{2B}a^\dagger\\
			\sqrt{2B}a & -\varepsilon
		\end{pmatrix}
		\begin{pmatrix}
			\psi_A\\
			\psi_B
		\end{pmatrix}
		=\begin{pmatrix}
			0\\
			0
		\end{pmatrix}
	\end{equation}
	giving rises to two coupled equations
	\begin{align}
		\sqrt{2B}a^\dagger \psi_B= \varepsilon \psi_A, \qquad
		\sqrt{2B}a \psi_A= \varepsilon \psi_B
	\end{align}
and they can be mapped to get a second differential equation having the Hermite polynomial as a solution. Then, as a result, one finds the eigenspinors
%	\cite{14}
	\begin{equation}\label{7}
		\Psi_{n,k_y}(x,y)=
		\begin{pmatrix}
			H_n \left(X\right)\\
			s H_{n-1} \left(X\right)
		\end{pmatrix} e^{-\frac{X^2}{2}}\ e^{ik_yy}
	\end{equation}
	associated to the quantized 
	eigenvalues
	\begin{align}\label{1111}
		\varepsilon_n= s \sqrt{2Bn}, \qquad n\in \mathbb{N}
	\end{align}
where the dimensionless variable $X=\frac{x+k_yl_B^2}{l_B}$ and    the magnetic length $l_B=\frac{1}{\sqrt{B}}$ are introduced. Here 
the sign function is $ s=\pm1 $. 
	
	\section{Full energy spectrum}\label{LLs}

Because the Hamiltonian $H_1$ is periodic, 
the Floquet theory can be used to calculate the temporal  spinor , with $T$ denotes transpose.
	From the eigenvalue equation, we integrate to obtain
	\begin{align}
		\phi_{A,B}(t)  = e^{-i\int_{0}^{t}U_0\cos(\omega t')dt'}
		e^{-iEt}
	\end{align}
which can be decomposed  in Fourier series 
\begin{align}
		\phi_{A,B}(t)	 =\sum_{n=-\infty}^{+\infty}D^n_{A,B} e^{-in\omega t}e^{-iEt}
	\end{align}
where $D^{n}_{A,B}$ are time-independent Fourier coefficients and $E$ is the Floquet energy.
	In the first order approximation $\pm \frac{\omega}{2}$ \cite{5},
the eigenspinors of $H_1$  can be written as
	\begin{equation}\label{8}
		\phi(t)
		=\begin{pmatrix}
			D^+_Ae^{i\frac{wt}{2}}+D^-_Ae^{-i\frac{wt}{2}}\\
			D^+_Be^{i\frac{wt}{2}}+D^-_Be^{-i\frac{wt}{2}}
		\end{pmatrix}\ e^{-iEt}.
	\end{equation}

	Finally, combining the solutions 
	\eqref{7} and \eqref{8} to build 
	$\Phi(x,y,t)=[\Phi_A(x,y,t),\Phi_B(x,y,t)]^T $ as eigenspinors 
the Hamiltonian \eqref{1}. Consequently, we get
	\begin{align}\label{9}
		\Phi_{n,k_y}(x,y,t)=
		\begin{pmatrix}
			H_n  
			\left[D^+_Ae^{i\frac{wt}{2}}+D^-_Ae^{-i\frac{wt}{2}}\right]\\
			s H_{n-1} \left[D^+_Be^{i\frac{wt}{2}}+D^-_Be^{-i\frac{wt}{2}}\right]
		\end{pmatrix} e^{-\frac{X^2}{2}}\ e^{ik_yy} \ e^{-iEt}.
	\end{align}	
	To determine the full energy spectrum we use the eigenvalue equation satisfied by the Hamiltonian \eqref{1} and  the  eigenspinors  \eqref{9}. As a result, we end up with 
	\begin{align}%\label{11}
		&		U_0\cos(\omega t)\Phi_A+\sqrt{2B}a^\dagger\Phi_B=i\frac{\partial}{\partial t}\Phi_A\label{10}\\
		&
			\sqrt{2B}a\Phi_A+U_0\cos(\omega t)\Phi_B=	i\frac{\partial}{\partial t}\Phi_B
	\end{align}
	and by  substituting the components of \eqref{9} we find
	%\begin{widetext}
		\begin{align}\label{12}
			& 
			U_0 \cos(\omega t)\left(D^+_Ae^{i\frac{wt}{2}}+D^-_Ae^{-i\frac{wt}{2}}\right)+
			\varepsilon_n\left(D^+_Be^{i\frac{wt}{2}}+D^-_Be^{-i\frac{wt}{2}}\right)= 
			E_-D^+_A e^{i\frac{wt}{2}}+E_+ D^-_A e^{-i\frac{wt}{2}}
			\\
			&
			U_0 \cos(\omega t)\left(D^+_Be^{i\frac{wt}{2}}+D^-_Be^{-i\frac{wt}{2}}\right)
			+ \varepsilon_n\left(D^+_Ae^{i\frac{wt}{2}}+D^-_Ae^{-i\frac{wt}{2}}\right)=   E_-D^+_B e^{i\frac{wt}{2}}+E_+D^-_B e^{-i\frac{wt}{2}}
			\label{13}
		\end{align}
%	\end{widetext}
	where we have set $E_\pm= E\pm \frac{w}{2}$. After some algebras and  
	within
	the 
	fast-oscillating terms  in the resonance approximation,
	we can neglect the term $e^{\pm i\frac{3\omega t}{2}}$ \cite{5}
to write 
	(\ref{12}-\ref{13}) as 
%	\begin{widetext}
		\begin{align}
			&\frac{U_0}{2}\left(D^+_Ae^{-i\frac{wt}{2}}+D^-_Ae^{i\frac{wt}{2}}\right)+\varepsilon_n\left(D^+_Be^{i\frac{wt}{2}}+D^-_Be^{-i\frac{wt}{2}}\right)=E_-D^+_Ae^{i\frac{wt}{2}}+E_+D^-_Ae^{-i\frac{wt}{2}} \\
			& \varepsilon_n	\left(D^+_Ae^{i\frac{wt}{2}}+D^-_Ae^{-i\frac{wt}{2}}\right)+\frac{U_0}{2}\left(D^+_Be^{-i\frac{wt}{2}}+D^-_Be^{i\frac{wt}{2}}\right)=E_- D^+_B e^{i\frac{wt}{2}}+E_+ D^-_B e^{-i\frac{wt}{2}}
		\end{align}
	%\end{widetext}
which can be cast in the matrix form
	\begin{align}\label{22}
		\begin{pmatrix}
			E_{-}&-\frac{U_0}{2}&-\varepsilon_n&0\\
			-\frac{U_0}{2}&E_{+}&0&-\varepsilon_n\\
			-\varepsilon_n&0& E_{-}&-\frac{U_0}{2}\\
			0&-\varepsilon_n&- \frac{U_0}{2}& E_{+}
		\end{pmatrix}
		\begin{pmatrix}
			D^+_A\\	D^-_A\\	D^+_B\\	D^-_B
		\end{pmatrix}=
		\begin{pmatrix}
			0\\0\\0\\0
		\end{pmatrix}.
	\end{align}
Actually, the eigenvalues can be obtained by requiring that the determinant of the matrix be null. Thus, after computation, we end with 	
	\begin{equation}\label{21}
		E_n^{ss'}=\varepsilon_n+\frac{s'}{2}\sqrt{{U_0^2+w^2}}
	\end{equation}
where the Landau levels $ \varepsilon_n $ is given in \eqref{1111} and $s'=\pm 1$. Indeed, first  the term $\frac{1}{2}\sqrt{{U_0^2+w^2}}$ 
	contributes by shifting the Landau levels up and downs,  known  as subbands. This result is in agreement with previous published works 
	on the subject \cite{Sabeeh08,Mekkaoui2014}. Second, we recall that such a term is found to play the rule of an opening gap by irradiating graphene with a
	 laser field \cite{5}.

\begin{figure}[H]
		\centering
		\includegraphics[scale=0.65]{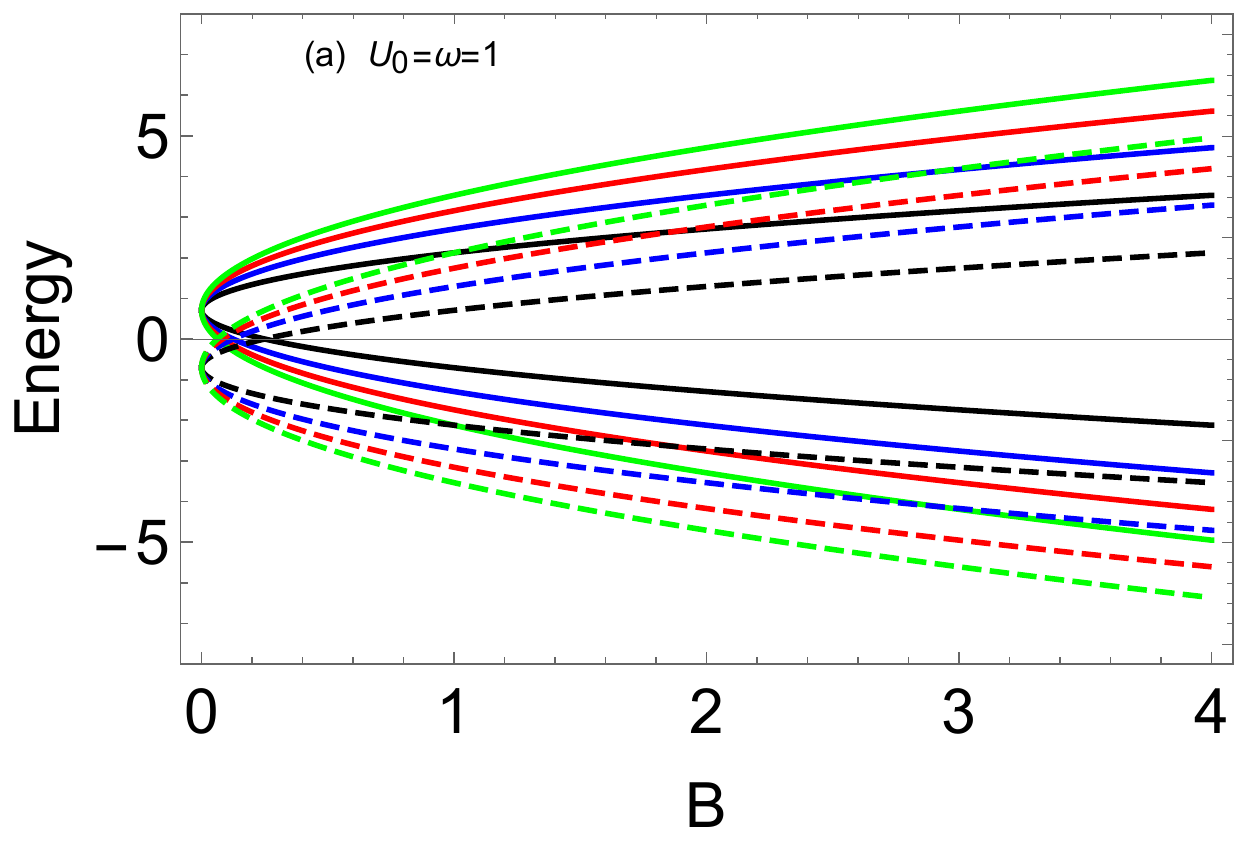}\ \ \
		\includegraphics[scale=0.65]{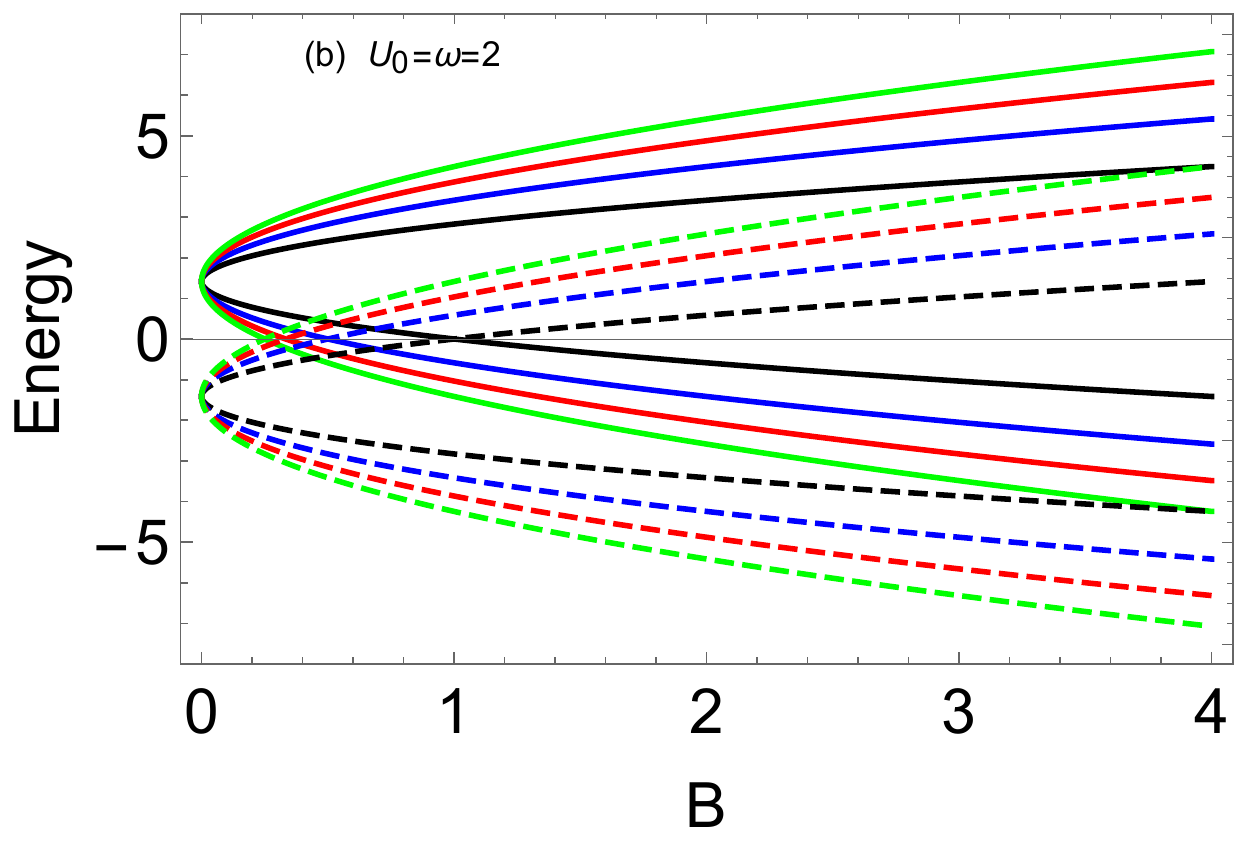}\\
		\includegraphics[scale=0.65]{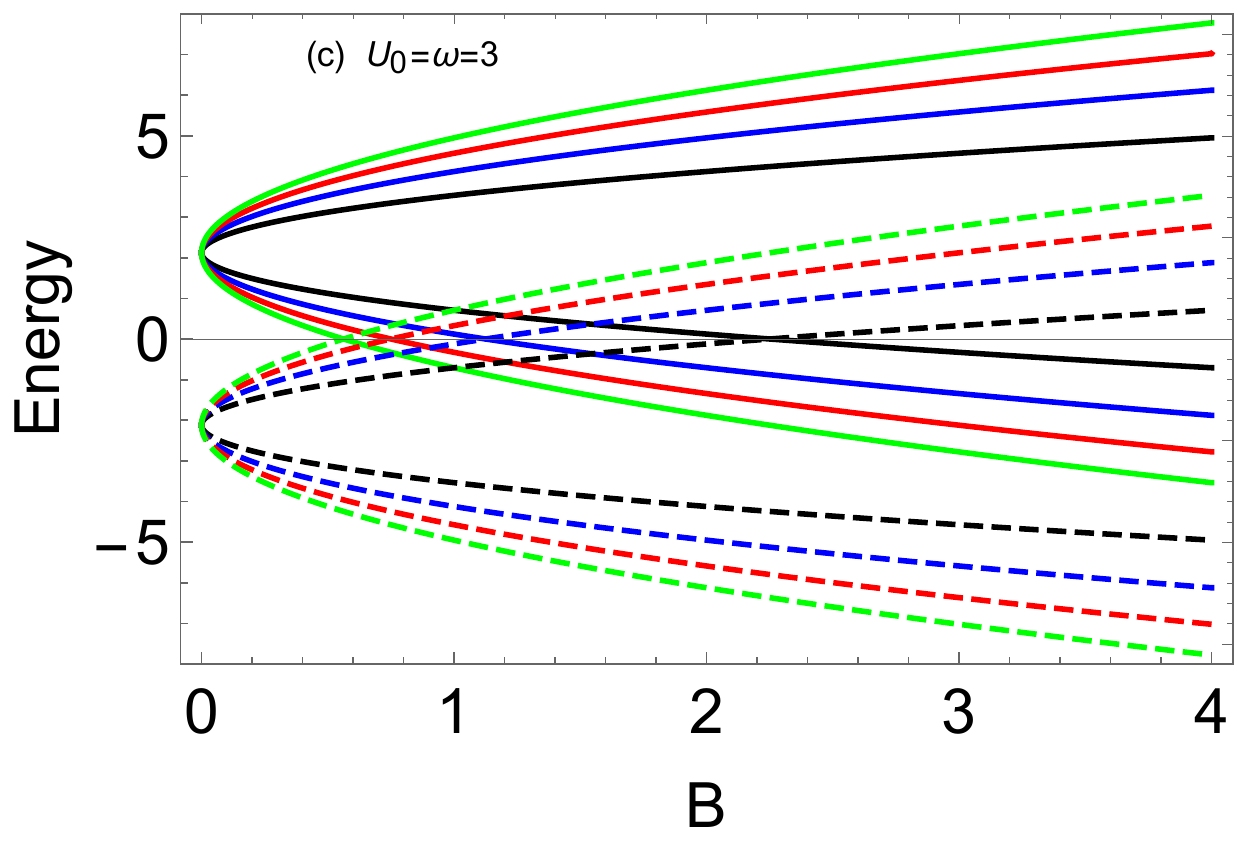}
		\ \ \
		\includegraphics[scale=0.65]{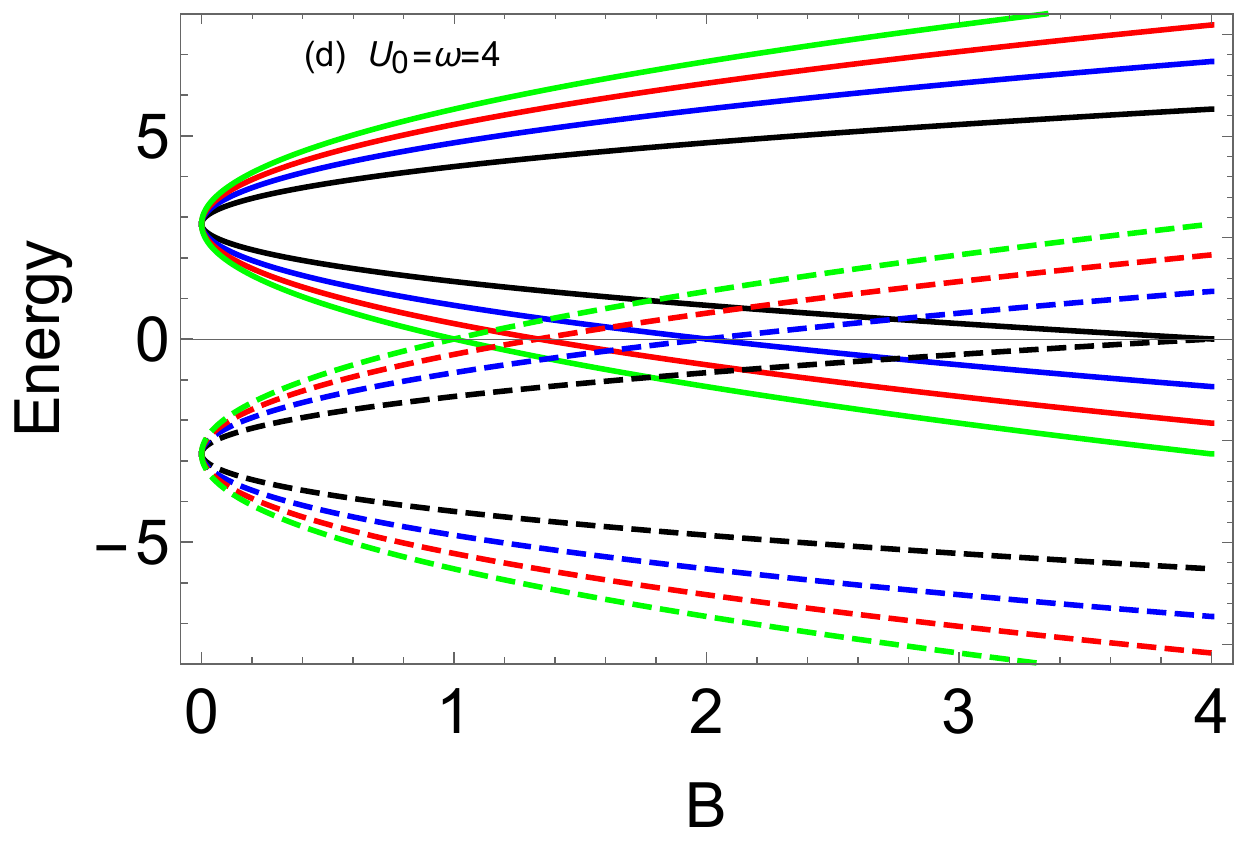}
		\caption{(color online) The energy spectrum as a function of the magnetic field $B$ with $U_0=\omega=1$ (a), 2 (b), 3 (c) and 4 (d). Four quantum numbers are considered  $ n=0 $	(black line), $ 1 $	(blue line), $ 2 $	(red line), $ 3 $	(green line). Continuous lines for $s'=1$ and dashed ones for $s'=-1$.}
		\label{fig1}
	\end{figure}

	We numerically analyze the effects of a magnetic field and an oscillating potential on Dirac fermions in graphene under various conditions.  In Fig. \ref{fig1}, we plot the energy $ E^{ss'}_n $ as a function of the magnetic field $B$ for
{four} quantum numbers $n$. We choose the parameters describing  the potential as $ U_0=\omega=1$ 
(a), 2 (b), 3 (c) and 4 (d). It is clearly seen  that the symmetry holds between different energy bands, that is  $ E^{++}_n= -  E^{--}_n$ and $ E^{-+}_n= -  E^{+-}_n$. Crossing points between different energy bands are observed as a result of the second term in  $ E^{ss'}_n $. 
As long as $ U_0$ and $\omega $ increase, these points change their positions.
 As  a result, we observe that the Landau levels are strongly affected  because of the shifts appearing in Fig. \ref{fig1}, which  depend on the oscillating potential. It might be relevant to mention the variations between the two consecutive energies and also under the change of $ s' $. They are
\begin{align}
&	E^{ss'}_{n+1} -E^{ss'}_{n} = s\sqrt{2B}\left(\sqrt{n+1}-\sqrt{n}\right)\\
&{E^{s+}_{n} -E^{s-}_{n} =\sqrt{U^2_0+ \omega^2}}.
\end{align} 

{Fig. \ref{fig2} shows the energy $ E^{ss'}_n $ as function of the 
amplitude $U_0$	of the oscillating 
	potential   for $\omega=1$ and  four values of the quantum number $n$ with  the magnetic field $ B=0.5 $T (a), {$ 1 $T} (b),  $1.5 $T (c) and  $ 2.5 $T (d). As one can see, the amplitude acts by increasing or decreasing all energy bands. Also, we notice the emergence of different crossing points between energy bands. One more result is that  the magnetic field affects the behavior of energy by creating different shifts as shown in Figs. \ref{fig2}(a,b,c,d).

	\begin{figure}[H]
		\centering
		\includegraphics[scale=0.65]{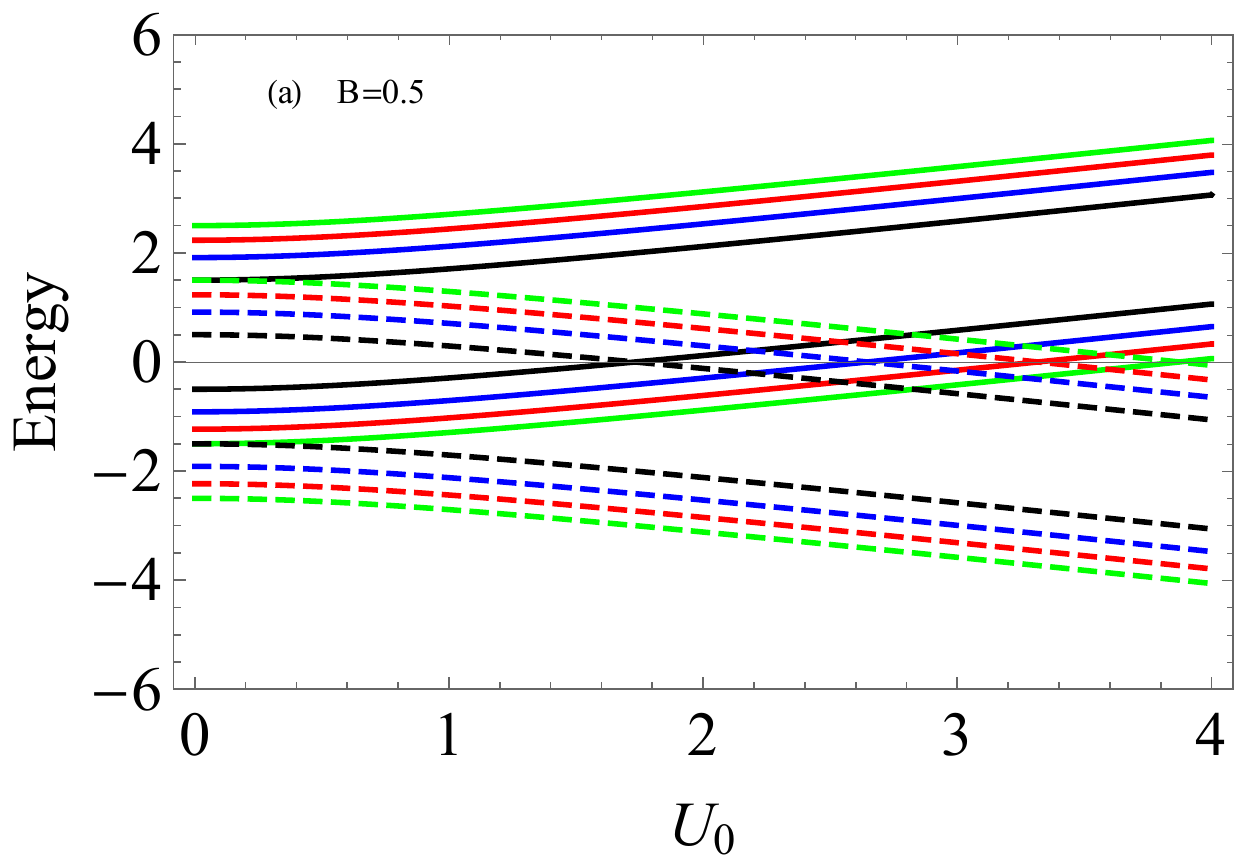}\ \ \
		\includegraphics[scale=0.65]{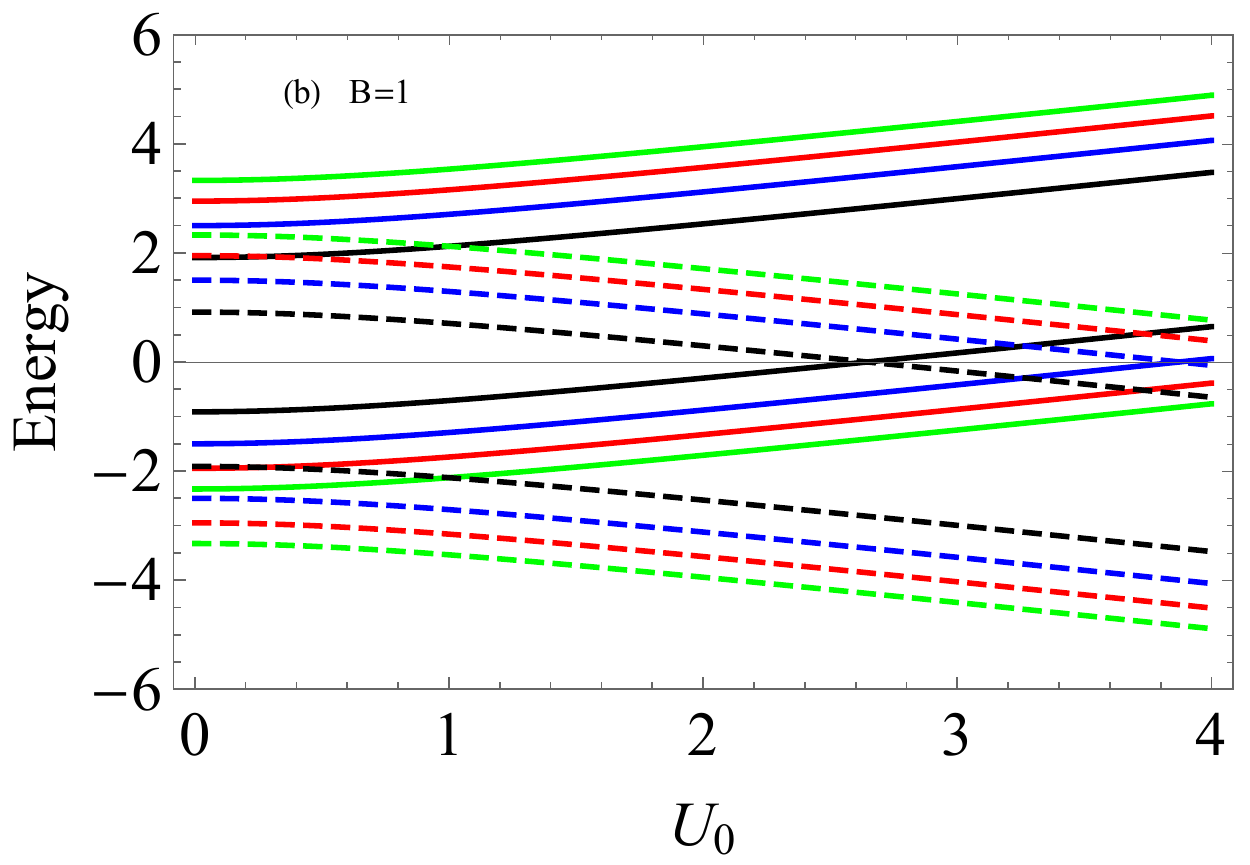}\\
		\includegraphics[scale=0.65]{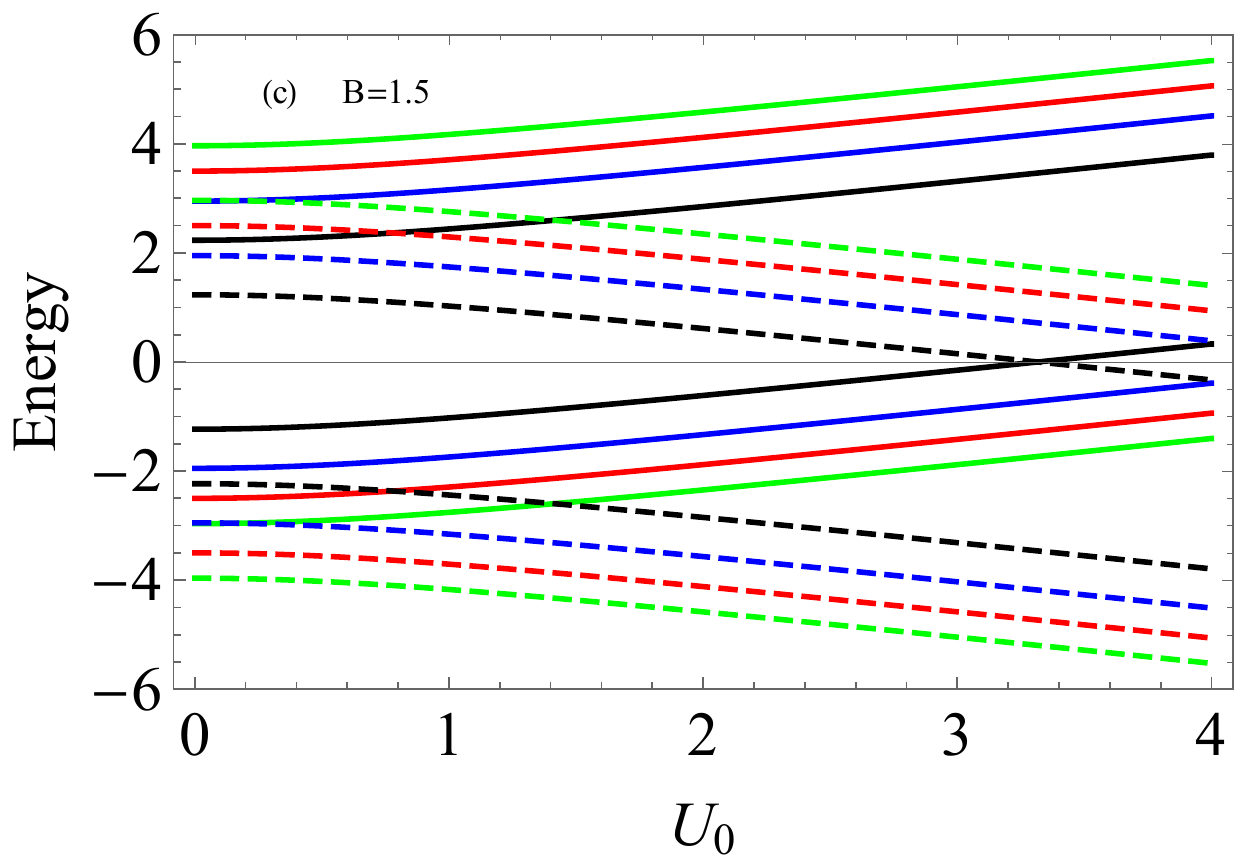}\ \ \
		\includegraphics[scale=0.65]{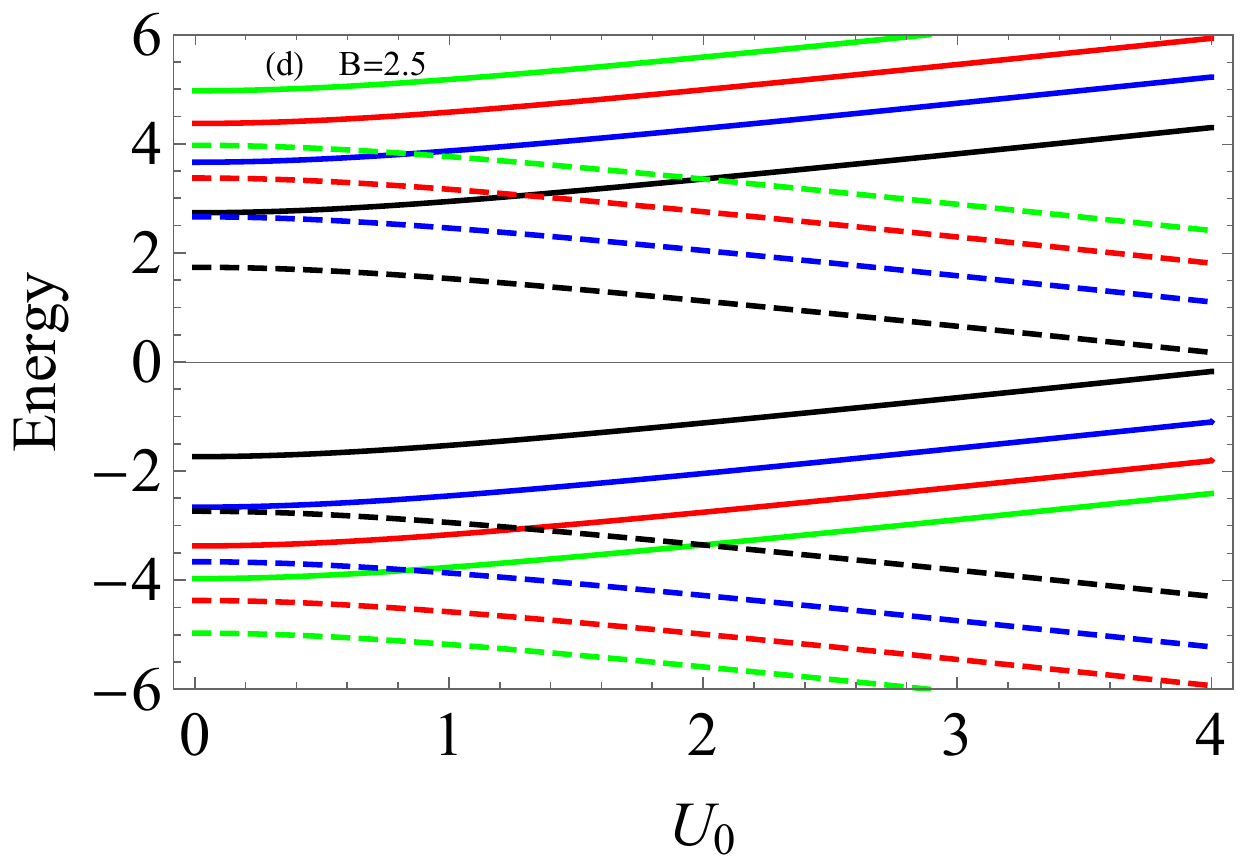}
		\caption{(color online) The energy spectrum as a function of the potential amplitude $U_0$ for $\omega=1$ with the quantum numbers $ n=1 $	(black line), $ 2 $	(blue line),
		$ 3 $	(red line), $ 4 $	(green line). Four values of magnetic fields are chosen  $ B=0.5 $T (a),   {$ 1 $T} (b),  $1.5 $T (c) and $2.5 $T (d). Continuous lines for $s'=1$ and dashed ones for $s'=-1$.}
		\label{fig2}
	\end{figure}

	\section{Current density}\label{CurrentD}
	
	We compute the current density $ \vec J $ through the
 continuity equation
	\begin{align}
	\frac{\partial}{\partial t}\rho+\vec \nabla\cdot  \vec J=0	
	\end{align}
such that the charge density is $\rho$ =$|\Psi|^2$. As for our system of  eigenspinors $ \Phi_{n,k_y}(x,y,t) $ \eqref{9}, we show that $ \vec J $ takes the following form
	\begin{align}
		\vec J=\Phi^*_{n,k_y}\vec \sigma\Phi_{n,k_y}
	\end{align}
and therefore we end up with  the two components
	take the forms
	\begin{align}\label{27}
		&	J_x^{(n)}=s\left[C_1+C_2\cos(wt)\right]H_n(X)H_{n-1}(X) \ e^{-X^2}\\
		& \label{26}
		J_y^{(n)}=sC\sin\left(wt\right)
		H_n(X)H_{n-1}(X)\ e^{-X^2}
	\end{align}
where we have defined
\begin{align}
&	C=2\left(D_A^+D_B^--D_A^-D_B^+\right)\\
&	C_1=2\left(D_A^+D_B^++D_A^-D_B^-\right)\\ &C_2=2\left(D_A^+D_B^-+D_A^-D_B^+\right).
\end{align}
For numerical use, it is convenient 
to rearrange the two components as 
	\begin{align}\label{27}
		&	J_x^{(n)}=J^n_{0x}+J^n_{1x}	\\
		& \label{26}
		J_y^{(n)}=J^n_{0y}+J^n_{1y}
	\end{align}
and here $ J^n_{0y}=0 $.
As one can notice,   
	the current densities generated in the $x$ and $y$-directions show two corrections, which are time dependent, i.e. $ J^n_{1x}$ and $ J^n_{1y} $. These are  analogous to those obtained by analyzing the Josephson current in graphene under periodic potential in position and time \cite{1900,11}.

	For the numerical implementation we focus only on the two quantum numbers $ n=1,2 $
	and  the Fourier coefficients $D^\pm_{A,B}=\frac{U_0}{2}$. The resulted   current densities along the $x$-direction are
	\begin{align}
		&J^{(1)}_{1x}=\pm 2U^2_0\cos(\omega t)Xe^{-X^2}\label{jx11}\\
		&J^{(2)}_{1x}=\pm 2U^2_0\cos(\omega t)X(4X^2-2)e^{-X^2}\label{jx12}
	\end{align}

	\begin{figure}[H]
		\centering
		\includegraphics[scale=0.6]{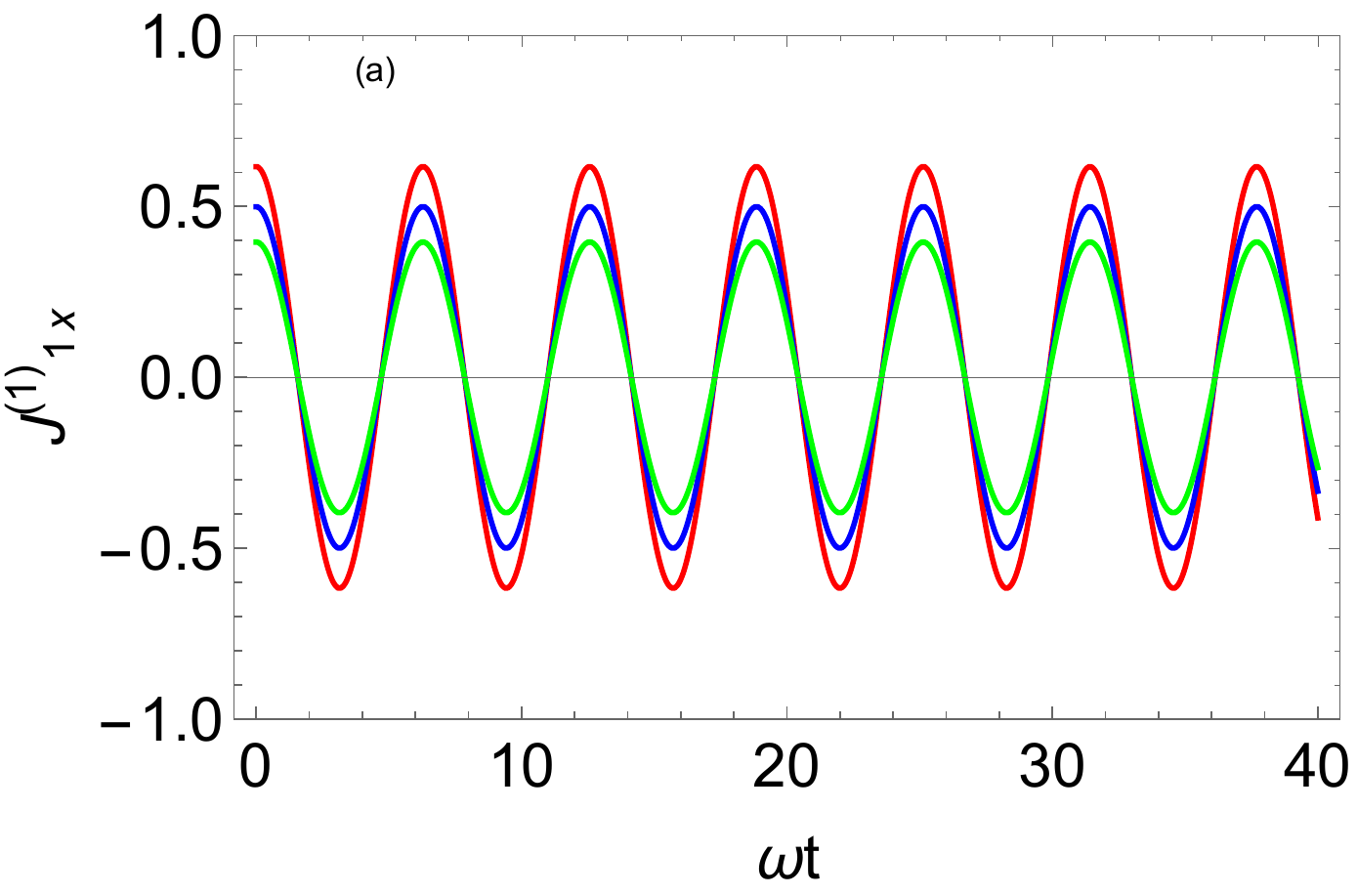}\ \ \
		\includegraphics[scale=0.6]{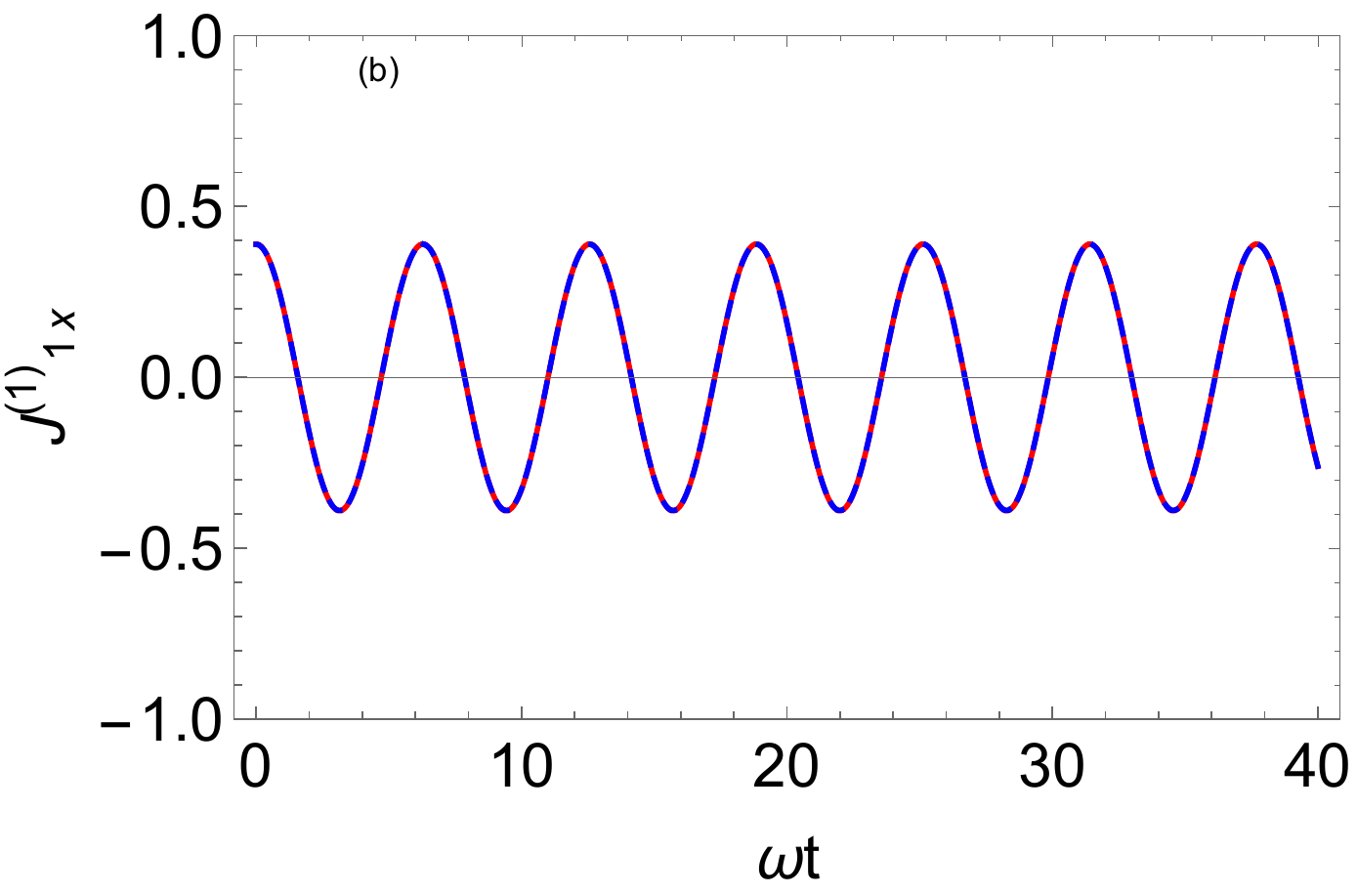}\\
		\includegraphics[scale=0.6]{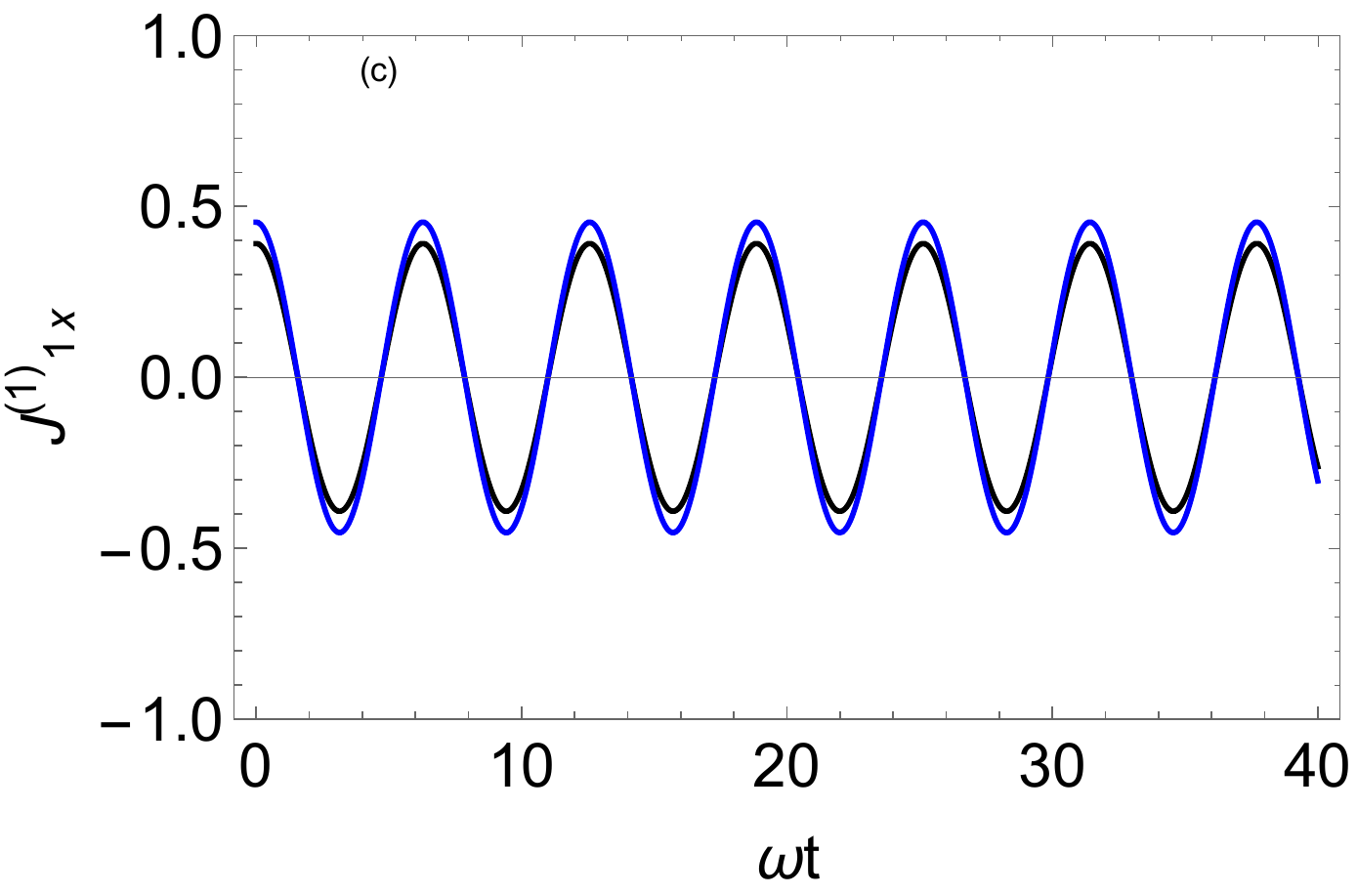}\ \ \
		\includegraphics[scale=0.6]{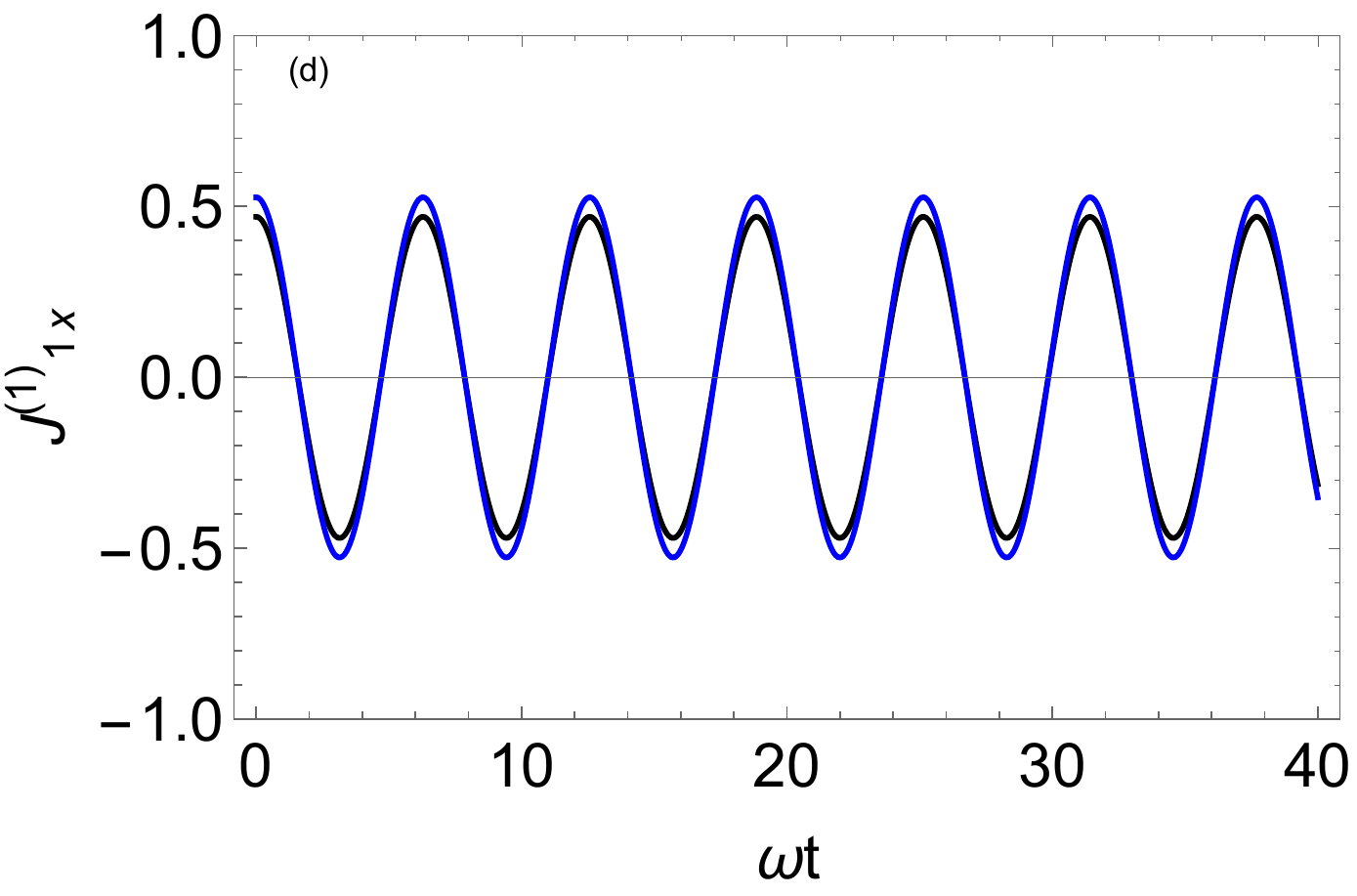}
	\caption{(color online) The current density $J^{(1)}_{1x}$ ($ n=1 $) as a function of $\omega t$ for $k=0.1$, $U_0=1$ and $x=0.1$. We choose  different values of magnetic field. {(a): $B=0.1$T (red line), $0.2$T (blue line), $0.6$T (green line).
		(b): $B=0.7$T (red line), $1.4$T (blue dashed). (c): $B=1.5$T (black line), $3.5$T (blue line).  (d): $B=4$T (black line), $6$T (blue line).}}
		\label{fig7}
	\end{figure}

{Fig. \ref{fig7}
	represents the current density $ J^{(1)}_{1x} $ ($ n=1 $) as a function of $\omega t$ under the choice of some values of the magnetic field $ B $. 
It is clearly seen  from  \eqref{jx11} that $ J^{(1)}_{1x} $ is a sinusoidal function in terms of $\omega t$ but with an amplitude of two contributions coming from $U_0$ and $ Xe^{-X^2} $.
	As a result, we see that there are three regimes $ B $ showing different behaviors. Indeed, we observe that from $0.1$T up to $0.6$T, $ B $ acts by decreasing the amplitude of oscillations (Fig. \ref{fig7}a) and  it 
becomes insensitive to $B $ regardless of the value taken (Fig. \ref{fig7}b)
	form {$0.7$T to $1.4$T. However, from $B=1.5$T}, we observe that 
the amplitude of oscillations starts to increase again (Fig. \ref{fig7}c) for a strong field.	This tells us that the magnetic field can serve as a control to manipulate the current density for experimental use.

	\begin{figure}[H]
	\centering
	\includegraphics[scale=0.4]{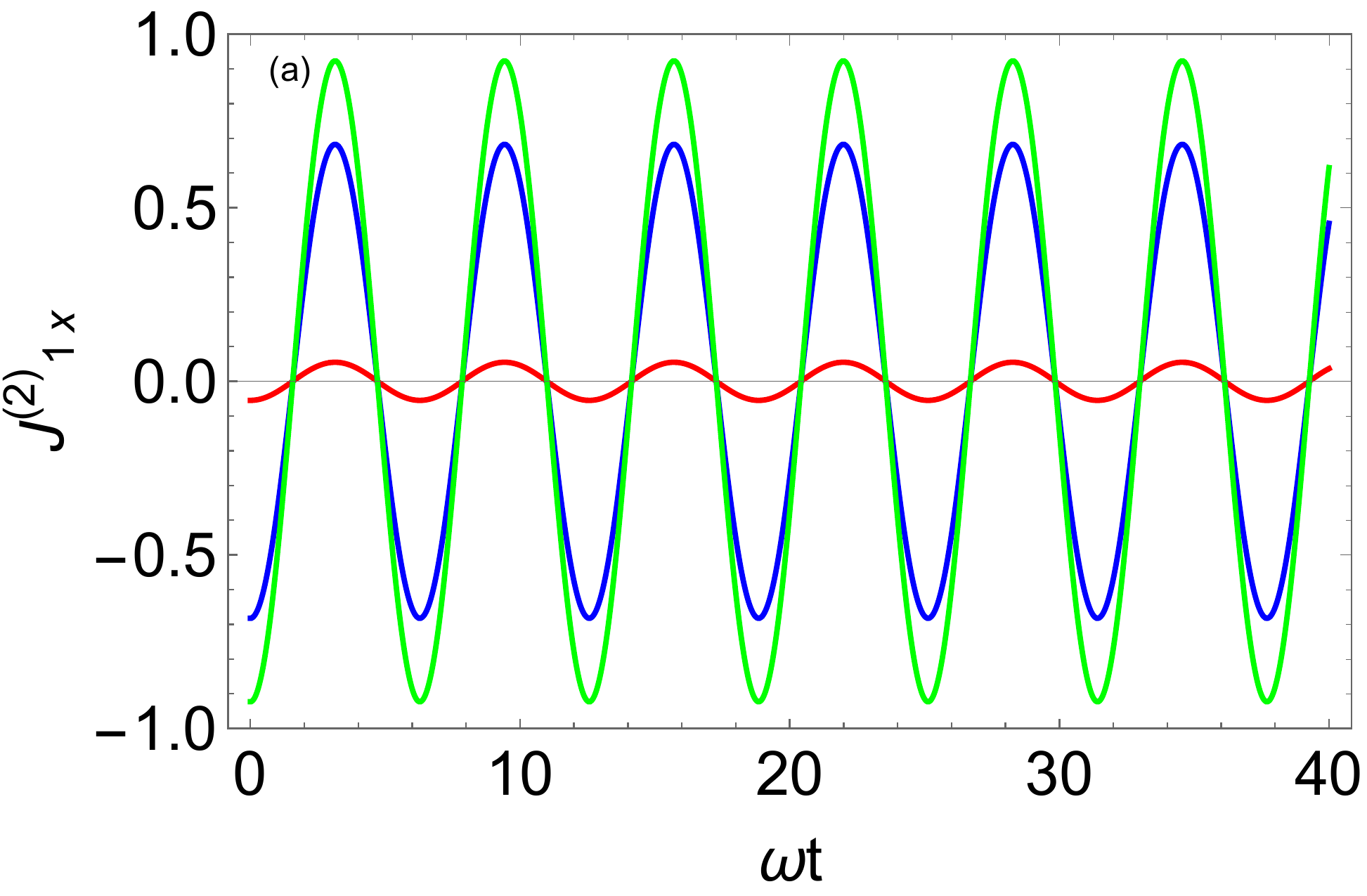}\ \ \ 
	\includegraphics[scale=0.4]{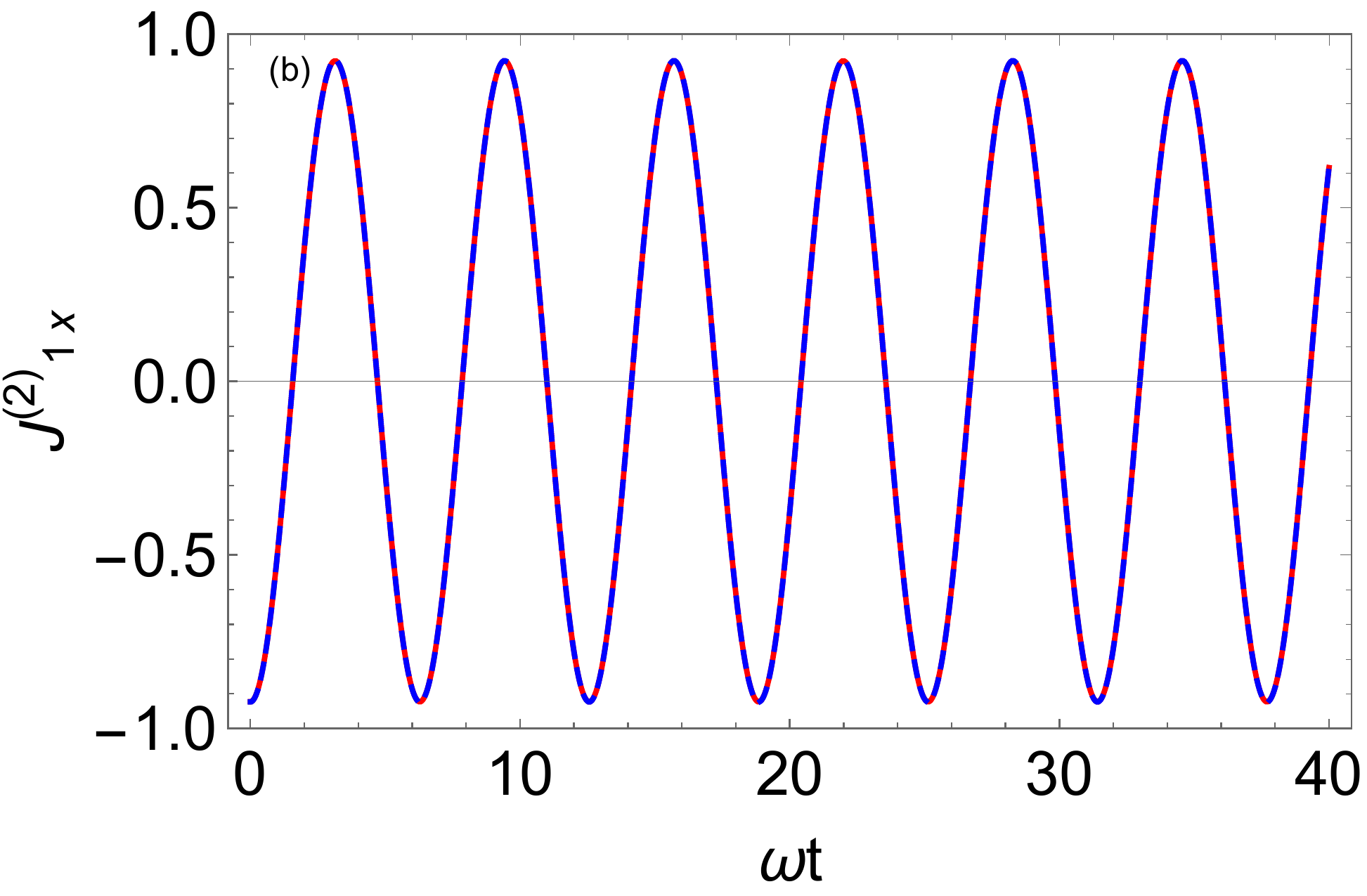}
	\includegraphics[scale=0.4]{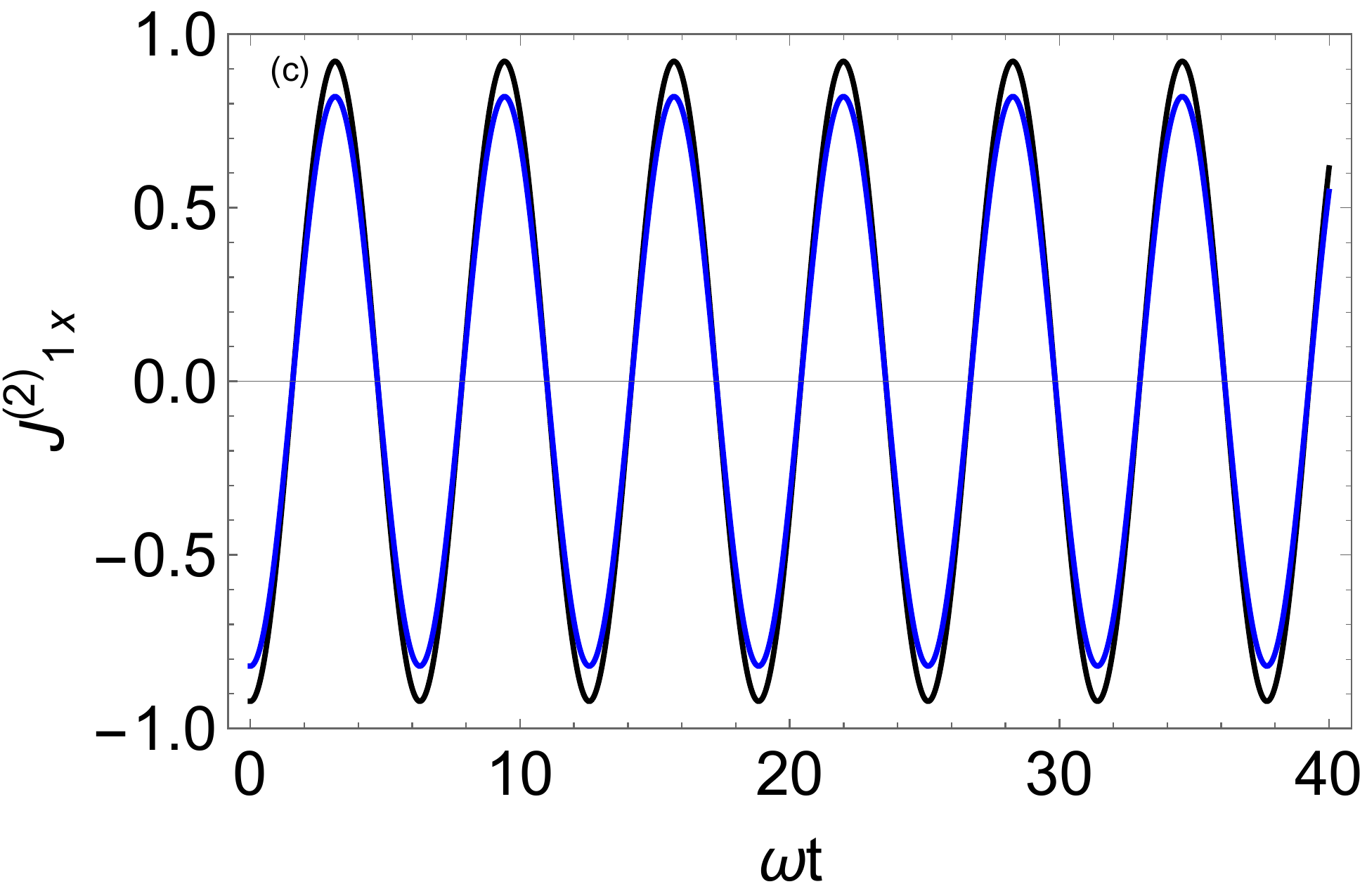}\ \ \
		\includegraphics[scale=0.4]{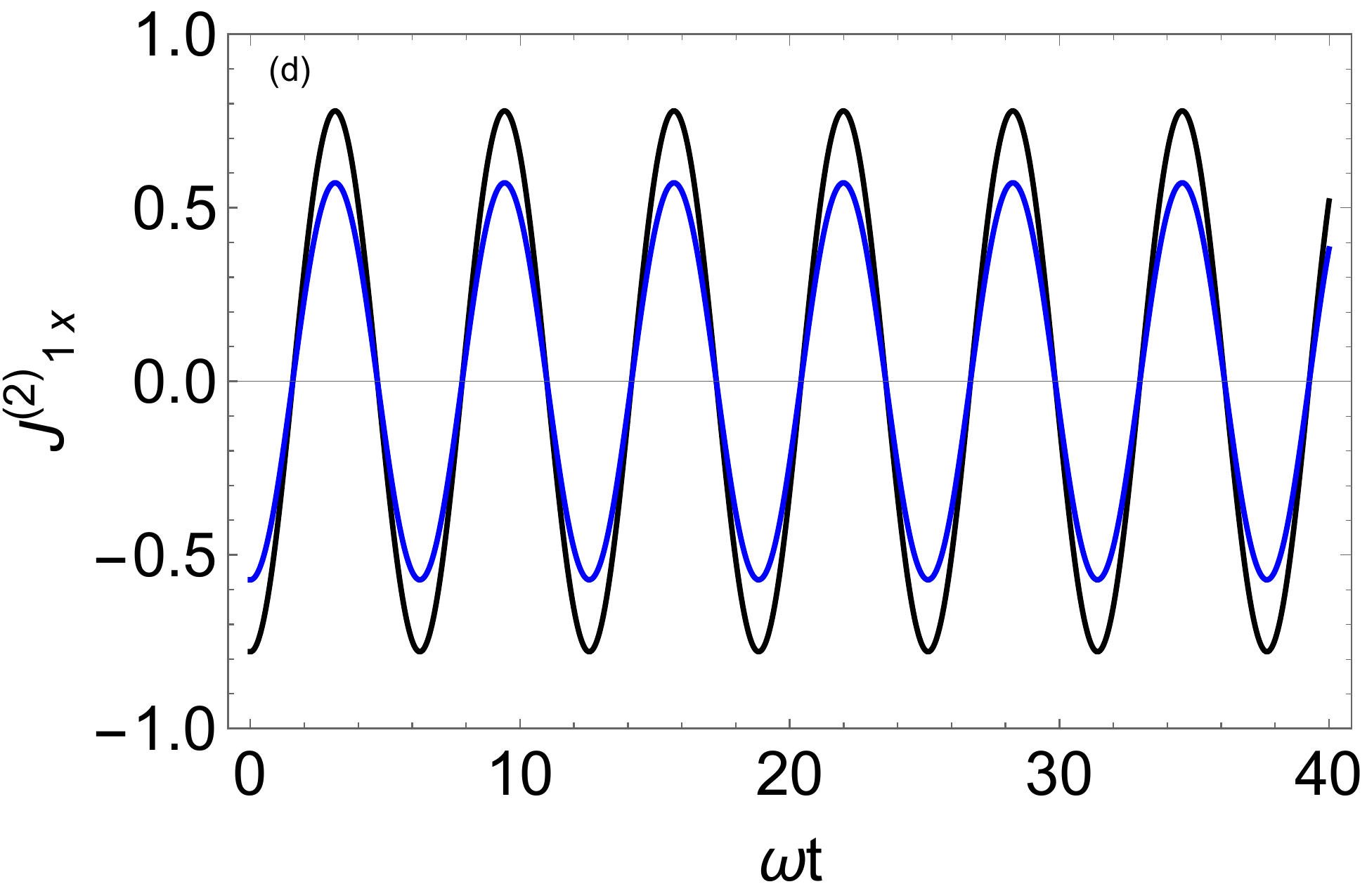}
\caption{(color online) The current density $J^{(2)}_{1x}$ ($ n=2 $) as a function of $\omega t$ for $k_y=0.2$, $U_0=1$ and $x=0.2$. We choose  different values of magnetic field. (a): { $B=0.1$T (red line), $0.2$T (blue line), $0.7$T (green line). (b): $B=0.7$T (red line), $1.4$T (blue dashed). (c): $B=1.5$T (black line), $3.5$T (blue line). (d): $B=4$T (black line), $6$T (blue line).} }
	\label{fig77}
\end{figure}

	\begin{figure}[H]
	\centering
	\includegraphics[scale=0.5]{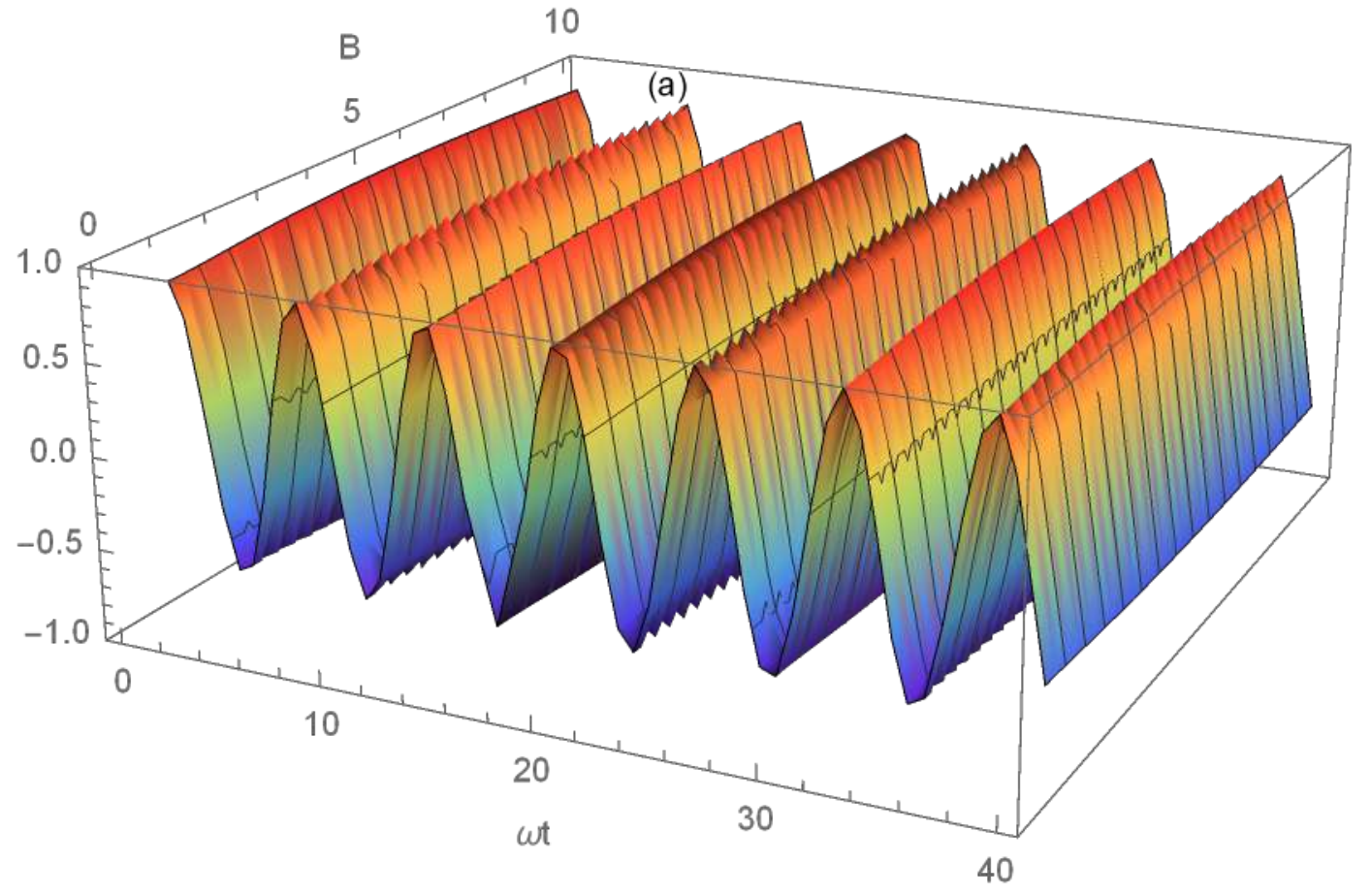}\ \ \ \ \ \
	\includegraphics[scale=0.5]{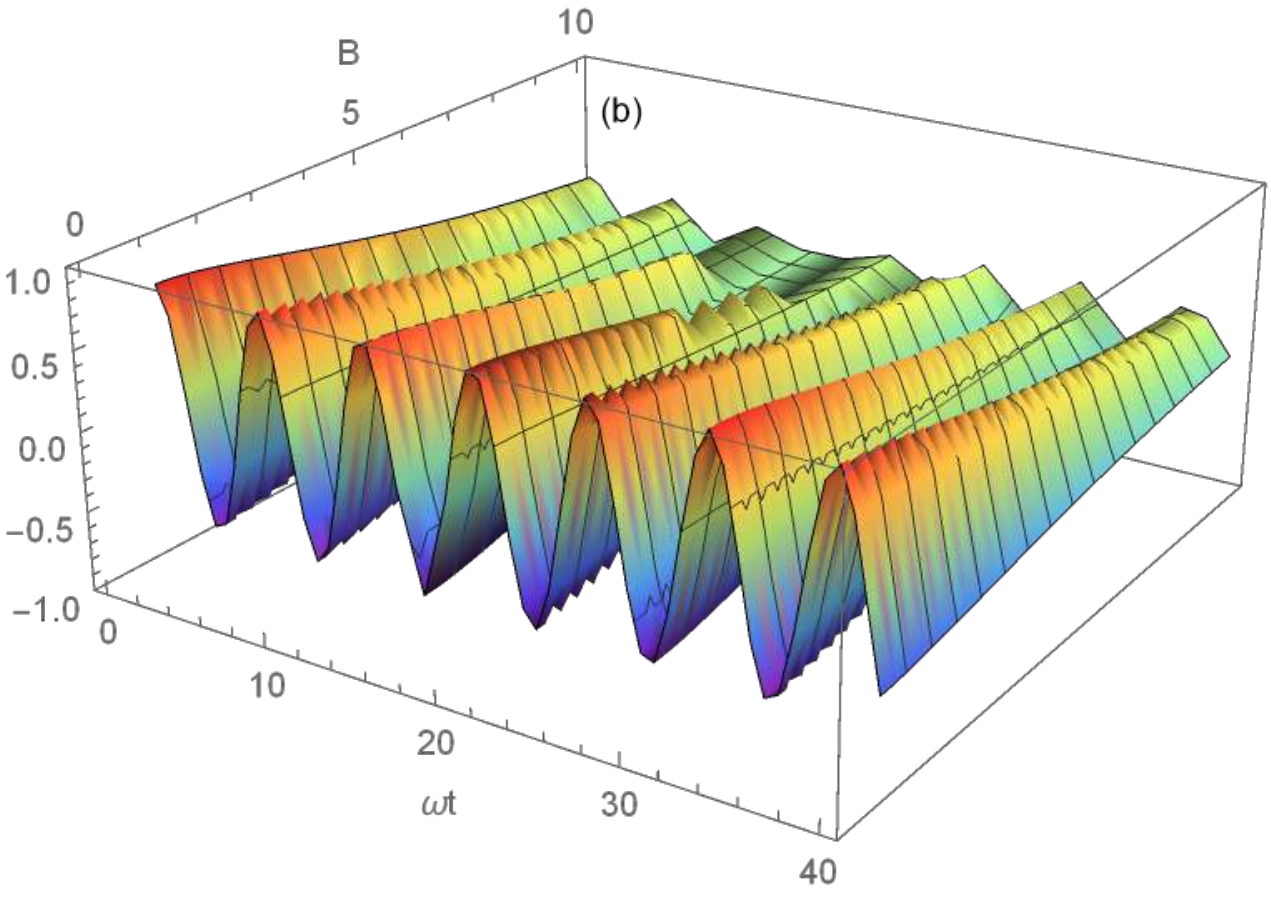}\\
	\includegraphics[scale=0.5]{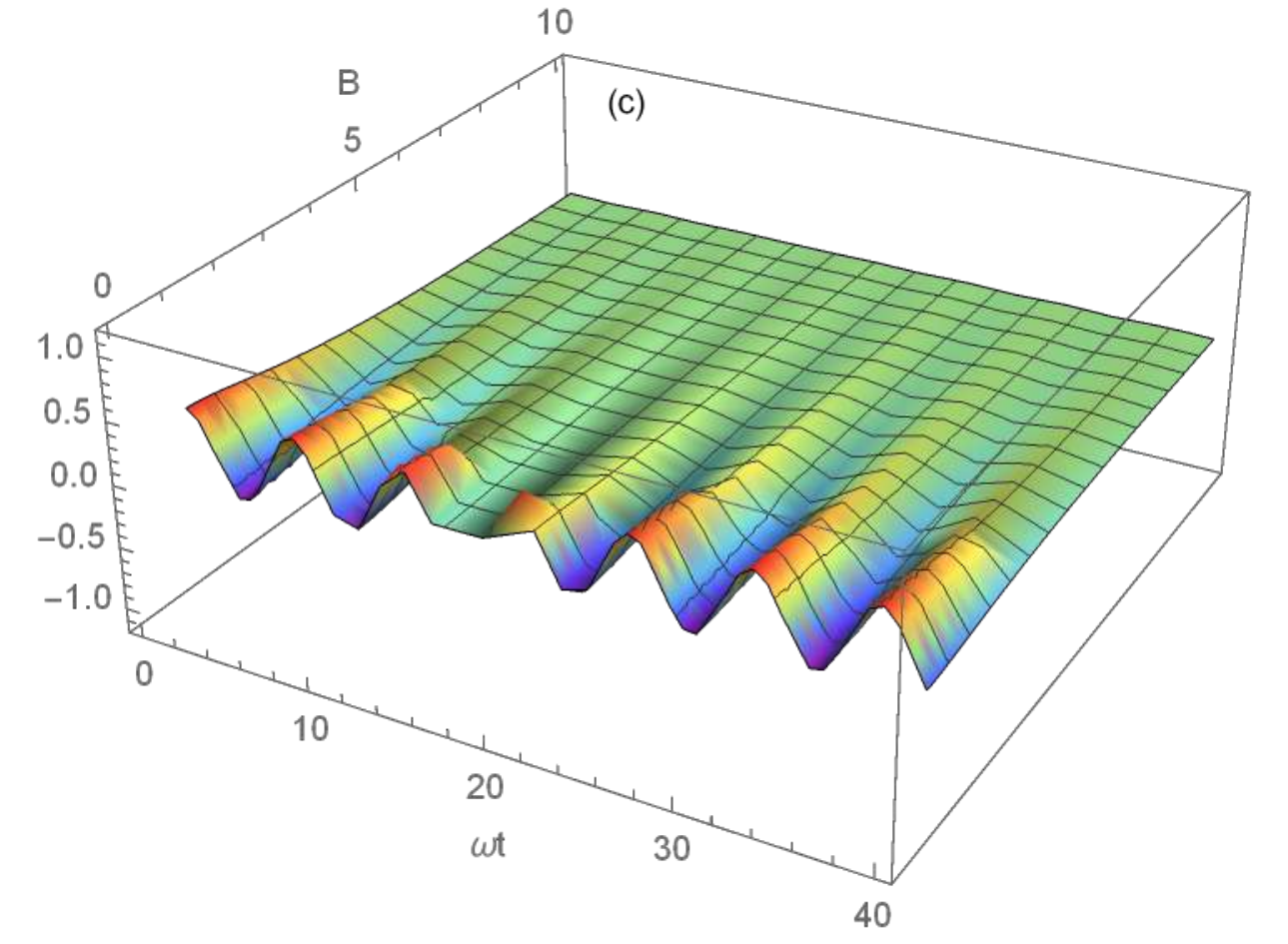}\ \ \ \ \ \
	\includegraphics[scale=0.5]{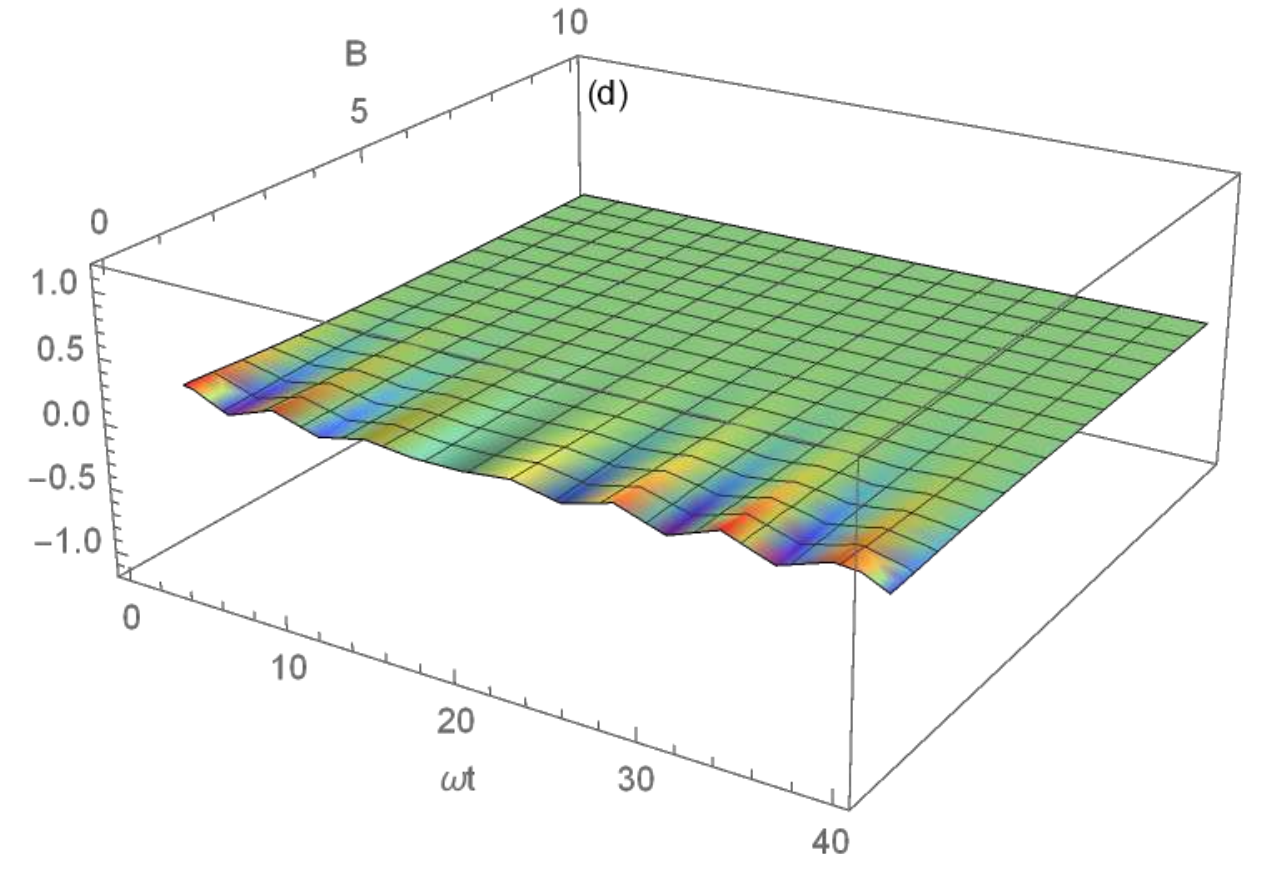}
	\caption{(color online) Density plot of the current density $J^{(1)}_{1x}$ as a function of the magnetic field $B$ and $\omega t$ for $U_0=1$ with four values of $k_y=x$ (a): $0.25$, (b): $0.5$, (c): $=0.75$ {and (d): $=1$.}}
	\label{fig9}
\end{figure}

In Fig. \ref{fig77}, we plot the  current density $J^{(2)}_{1x}$ ($ n=2 $) as a function of $\omega t$ and choose some values of the magnetic field $B$. We also have a sinusoidal function with the  same period but different amplitudes depending on $B$. As before,  the behavior of $J^{(2)}_{1x}$ can be divided into three zones according to the value taken by $B$. Indeed, the amplitude of oscillations { increases  
from  $0.1$T up to $0.7$T as shown in Fig. \ref{fig77}a	
	and remains constant from $0.7$T up to $1.4$T in Fig. \ref{fig77}b.
	However, as seen in 
Fig. \ref{fig77}c,	
	the amplitudes decrease despite increasing  the field to 
$1.5$T up to $3.5$T.} Just to illustrate, we observe that
for $B=0.2$, the
 amplitude is seven times greater than for $ B=0.1$. 

Fig.  \ref{fig9} represents the  current density $J^{(1)}_{1x}$
	 as a function of  the magnetic field and $\omega t$ by choosing {four} values of the wave vector and position such that $k_y=x$. As for Fig.  \ref{fig9}a with  $k_y=x=0.25$, 
	 we observe that for each value of $B$, the current density is sinusoidal and for each half period $\omega =\frac{T}{2}$,  the peaks occur. It is clearly seen that for $k_y=x=0.5$ in Fig.  \ref{fig9}b  the  amplitude decreases as long as $B$ increases. In Fig.  \ref{fig9}c for
	 $k_y=x=0.75$ clearly shows the attenuation of $J^{(1)}_{1x}$ with the magnetic field. More precisely, after a certain value of $B$, there is no current despite  increasing it.

	\begin{figure}[H]

	\centering
	\includegraphics[scale=0.55]{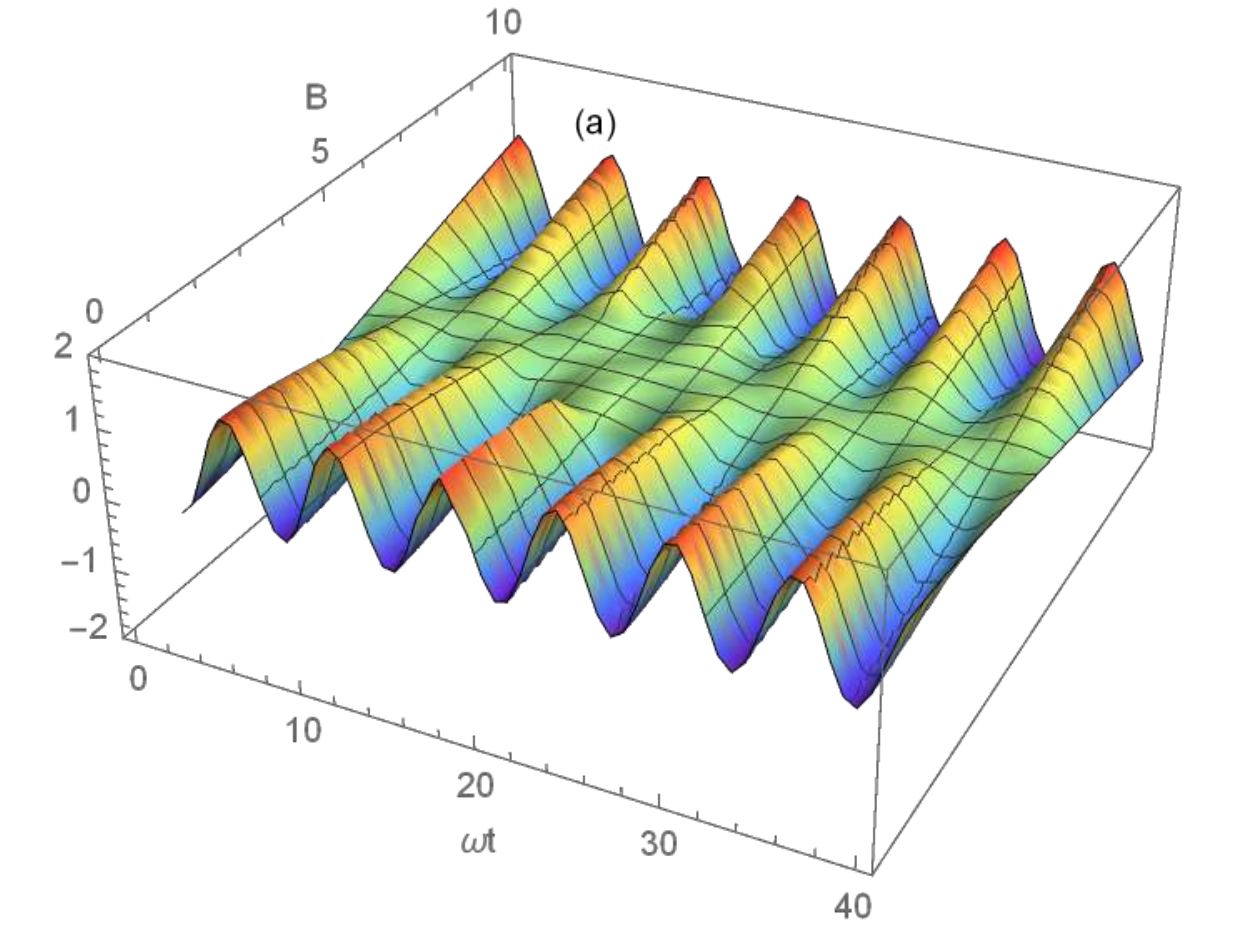}\ \ \ \ \ \
	\includegraphics[scale=0.55]{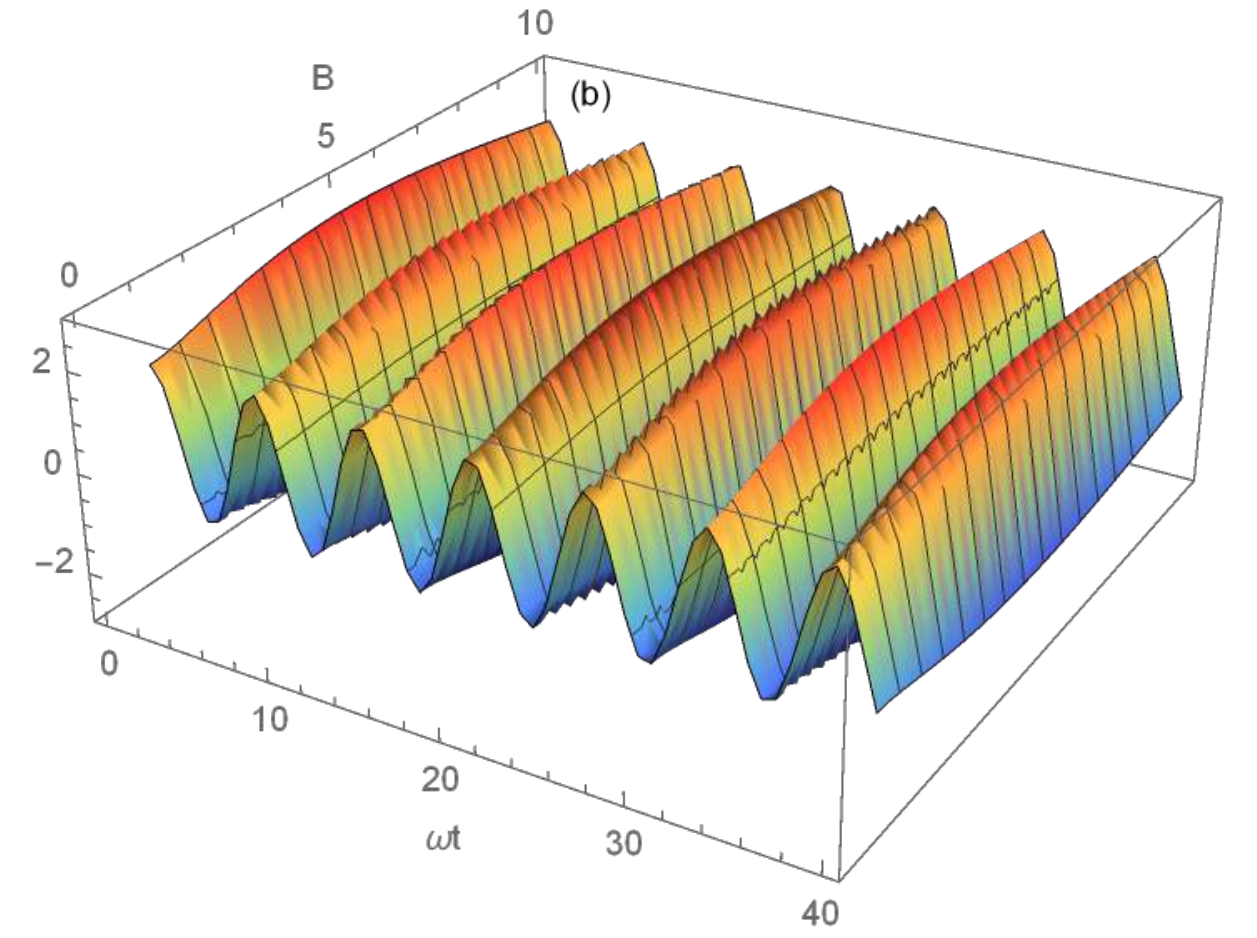}\\
	\includegraphics[scale=0.55]{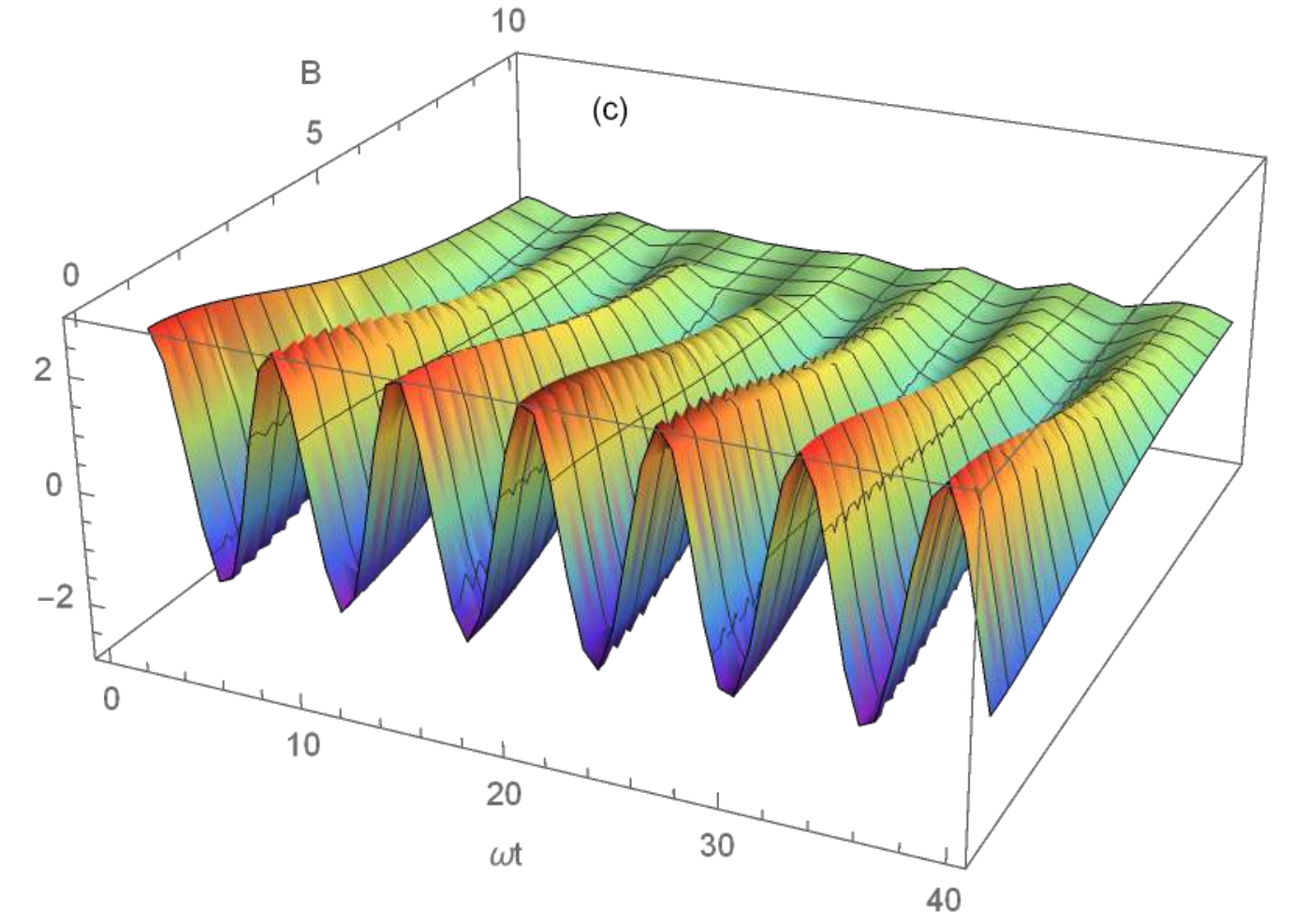}\ \ \ \ \ \
	\includegraphics[scale=0.55]{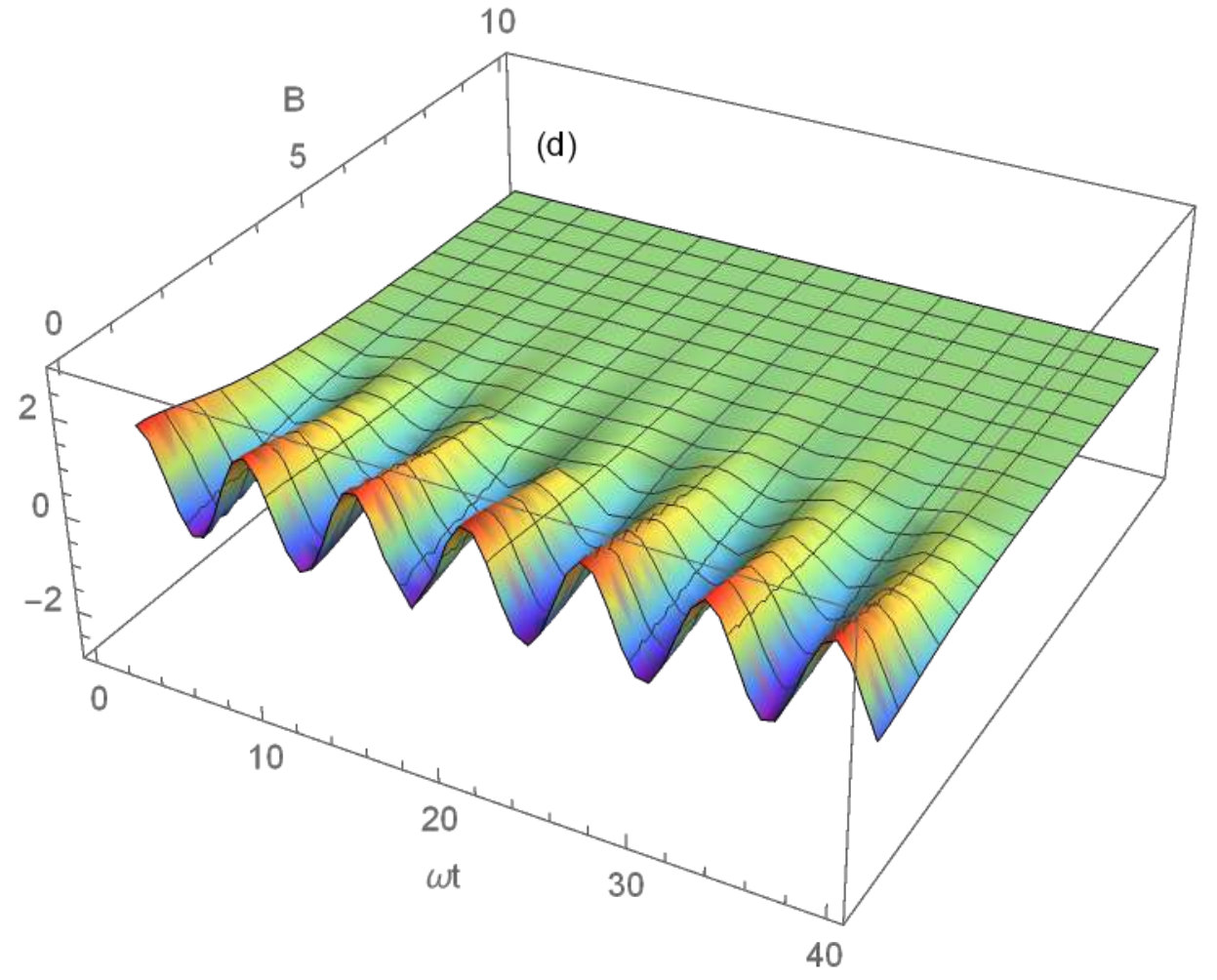}
	\caption{(color online) Density plot of the current density $J^{(2)}_{1x}$ as a function of $B$ and $\omega t$ for $U_0=1$ with four value of $k_y=x$ (a): $0.25$, (b):~$0.5$ and (c): $0.75$ and (d): $=1$.}
	\label{fig11}
\end{figure}

Fig. \ref{fig11} represents the current density  $J^{(2)}_{1x}$ as a function of  $B$ and $\omega t$ for three values of $k_y=x $. As for 
$k_y=x=0.25$ we notice that Fig. \ref{fig11}a	
	shows that the curve is symmetrical with respect to an axis  perpendicular to the value $B=5$ because of the term $ (4X^2-2) $ appearing in \eqref{jx12}. One more thing is that
	 the amplitude of
$J^{(2)}_{1x}$	 
	 is small compared to that of $J^ {(1)}_{1x}$ as seen in Fig. \ref{fig9}a.
We also notice that far from the axis of symmetry, $J^{(2)}_{1x}$ exhibits  sinusoidal behavior with an increase in amplitude. For $k y=x=0.5$
Fig. \ref{fig11}b  shows that the amplitude of oscillations increases in the vicinity of  $B=5$ and decreases elsewhere, but  $J^{(2)}_{1x}$ is still a sinusoidal function. In
Fig. \ref{fig11}c  for $k_y=x=0.75$, we observe an attenuation of the current density when B increases, and then, after a certain value of B, there is no current. These results show evidence of the effect of the magnetic field on the current density of graphene subjected to an oscillating potential.

%%%%%%%%%%%%%%%%%%%%%%%%%%%%%%%%%%%%%
	\section{Conclusion}
%%%%%%%%%%%%%%%%%%%%%%%%%%%%%%%%%%%%%%

	We have studied the effect of a magnetic field on graphene in the presence of a temporal potential. Solving the Dirac equation, we have derived the exact solutions of the energy spectrum using the Floquet theory together with algebraic method. The eigenvalues are found as Landau levels added to extra subbands originated from the oscillating potential in the first approximation of the frequency $\omega$. It is showed that the energy exhibits a symmetry under the change of energy signs in addition to crossing points between different bands. Our results are an agreement with precious published works on the subject \cite{Sabeeh08,Mekkaoui2014}.
	
Subsequently, we have determined the current density in $x$ and $y$-directions. It consists of two separate parts  depending on the  position dimensionless $X$ and the oscillating potential. It was found that the current density  exhibits  extra terms that originated from oscillating potential, knowing that they could be interpreted as corrections to standard results. Our numerical analysis showed that  the current density oscillates with different amplitudes strongly dependent on the magnetic field $B$. In fact, we have found three sets of $B$ generating various amplitudes of oscillation. More precisely, the amplitude can decrease, increase, or remain constant according to the value taken by $B$.	
Furthermore, the wave vector $k_y$ and space position $x$ are discovered to be important in the density plot of the current density in terms of $B$ and the argument $\omega t$. As a result, they act by decreasing the current density to a very low level.

\end{document}